\input lanlmac
\def\href#1#2{{#2}}
\def\hhref#1{{#1}}
\input epsf.tex

\overfullrule=0mm

\newcount\figno
\figno=0
\def\fig#1#2#3{
\par\begingroup\parindent=0pt\leftskip=1cm\rightskip=1cm\parindent=0pt
\baselineskip=11pt
\global\advance\figno by 1
\midinsert
\epsfxsize=#3
\centerline{\epsfbox{#2}}
\vskip 12pt
{\bf Fig.\ \the\figno:} #1\par
\endinsert\endgroup\par
}
\def\figlabel#1{\xdef#1{\the\figno}}
\def\encadremath#1{\vbox{\hrule\hbox{\vrule\kern8pt\vbox{\kern8pt
\hbox{$\displaystyle #1$}\kern8pt}
\kern8pt\vrule}\hrule}}


\magnification=\magstep1
\baselineskip=12pt
\hsize=6.3truein
\vsize=8.7truein
 at 8truept
 at 8truept
 at 10truept

\vbox{\hfill IPhT-t12/013}
\bigskip
\bigskip
\font\bigrm=cmr12 at 14pt \centerline{\bigrm More on the  O(n) model on random 
maps}
\medskip
\centerline{\bigrm via nested loops:}
\medskip
\centerline{\bigrm loops with bending energy}

\bigskip\bigskip

\centerline{G. Borot$^1$, J. Bouttier$^2$ and E. Guitter$^2$}
  \smallskip
  \centerline{$^1$ Section de Math\'ematiques}
  \centerline{Universit\'e de Gen\`eve}
  \centerline{2-4 rue du Li\`evre, Case postale 64, 1211 Gen\`eve 4, Suisse}
  \centerline{$^2$ Institut de Physique Th\'eorique}
  \centerline{CEA, IPhT, F-91191 Gif-sur-Yvette, France}
  \centerline{CNRS, URA 2306}
\centerline{\tt gaetan.borot@unige.ch}
\centerline{\tt jeremie.bouttier@cea.fr}
\centerline{\tt emmanuel.guitter@cea.fr}

  \bigskip

     \bigskip\bigskip

     \centerline{\bf Abstract}
     \smallskip
     {\narrower\noindent
We continue our investigation of the nested loop approach to the
$O(n)$ model on random maps, by extending it to the case where loops
may visit faces of arbitrary degree. This allows to express the
partition function of the $O(n)$ loop model as a specialization of the
multivariate generating function of maps with controlled face degrees,
where the face weights are determined by a fixed point condition. We
deduce a functional equation for the resolvent of the model, involving
some ring generating function describing the immediate vicinity of the
loops. When the ring generating function has a single pole, the model
is amenable to a full solution. Physically, such situation is realized
upon considering loops visiting triangles only and further weighting
these loops by some local bending energy. Our model interpolates
between the two previously solved cases of triangulations without
bending energy and quadrangulations with rigid loops. We analyze the
phase diagram of our model in details and derive in particular the
location of its non-generic critical points, which are in the
universality classes of the dense and dilute $O(n)$ model coupled to
2D quantum gravity.  Similar techniques are also used to solve a
twisting loop model on quadrangulations where loops are forced to make
turns within each visited square. Along the way, we revisit the
problem of maps with controlled, possibly unbounded, face degrees and
give combinatorial derivations of the one-cut lemma and of the
functional equation for the resolvent.
\par}

    \bigskip

\nref\TutteCPM{W.T. Tutte, {\it A Census of Planar Maps}, Canad. J. of Math.
{\bf 15} (1963) 249-271.}
\nref\LGMBuz{See for instance: J.-F. Le Gall and G. Miermont, {\it
Scaling limits of random trees and planar maps}, Lecture notes of the
Clay Mathematical Institute Summer School, Buzios (2010),
arXiv:1101.4856, and references therein.}
\nref\DGZ{See for instance: P. Di Francesco, P. Ginsparg and
J. Zinn--Justin, {\it 2D Gravity and Random Matrices},
Physics Reports {\bf 254} (1995) 1-131,	arXiv:hep-th/9306153, 
and references therein.}
\nref\DK{B. Duplantier and I. Kostov, {\it Conformal spectra of polymers on
a random surface}, Phys. Rev. Lett. {\bf 61} (1988) 1433-1437.}
\nref\Kost{I. Kostov, {\it $O(n)$ vector model on a planar random lattice:
spectrum of anomalous dimensions}, Mod. Phys. Lett. {\bf 4} (1989) 217-226.}
\nref\KostSta{I. Kostov and M. Staudacher, {\it Multicritical Phases of the 
O(n) Model on a Random Lattice}, Nucl. Phys. {\bf B384} (1992) 459-483,
arXiv:hep-th/9203030.}
\nref\EZJ{B. Eynard and J. Zinn--Justin, {\it The $O(n)$ model on a random
surface: critical points and large order behaviour}, Nucl. Phys. {\bf B386}
(1992) 558-591, arXiv:hep-th/9204082.}
\nref\EK{B. Eynard and C. Kristjansen, {\it Exact solution of the $O(n)$
model on a random lattice}, Nucl. Phys. {\bf B455} (1995) 577-618, 
arXiv:hep-th/9506193.}
\nref\EKmore{B. Eynard and C. Kristjansen, {\it More on the exact solution
of the $O(n)$ model on a random lattice and an investigation of the
case $|n|>2$}, Nucl. Phys. {\bf B466} (1996) 463-487,
arXiv:hep-th/9512052.}
\nref\BE{G. Borot and B. Eynard, {\it Enumeration of maps with self avoiding 
loops and the O(n) model on random lattices of all topologies}, 
J. Stat. Mech. (2011) P01010, arXiv:0910.5896.}
\nref\KPZ{V.G. Knizhnik, A.M. Polyakov and A.B. Zamolodchikov, {\it Fractal Structure of
2D Quantum Gravity}, Mod. Phys. Lett.
{\bf A3} (1988) 819-826; F. David, {\it Conformal Field Theories Coupled to 2D Gravity in the
Conformal Gauge}, Mod. Phys. Lett. {\bf A3} (1988) 1651-1656; J.
Distler and H. Kawai, {\it Conformal Field Theory and 2D Quantum Gravity}, 
Nucl. Phys. {\bf B321} (1989) 509-527.}
\nref\MierUniq{G. Miermont, {\it The Brownian map is the scaling limit
of uniform random plane quadrangulations}, arXiv:1104.1606.}
\nref\LGUniq{J.-F. Le Gall, {\it Uniqueness and universality of the
Brownian map}, arXiv:1105.4842.}
\nref\LGM{J.-F. Le Gall and G. Miermont, {\it Scaling limits of random planar 
maps with large faces}, Ann. Probab. {\bf 39(1)} (2011) 1-69, arXiv:0907.3262 
[math.PR].}
\nref\BBG{G. Borot, J. Bouttier and E. Guitter, {\it A recursive approach to the $O(n)$ model on random maps via nested loops}, J. Phys. A: Math. Theor. 45 (2012) 045002, arXiv:math-ph/1106.0153.}
\nref\GBThese{G.~Borot, PhD Thesis, Universit\'{e} d'Orsay (2011),
available at \hhref{http://tel.archives-ouvertes.fr/tel-00625776/en/}.}
\nref\Lorentz{P.~Di Francesco, E.~Guitter and C.~Kristjansen, {\it Integrable
2D Lorentzian Gravity and Random Walks}, Nucl. Phys. {\bf B567} (2000) 515-553,
arXiv:hep-th/9907084.}
\nref\CENSUS{J. Bouttier, P. Di Francesco and E. Guitter, {\it Census of planar
maps: from the one-matrix model solution to a combinatorial proof},
Nucl. Phys. {\bf B645}[PM] (2002) 477-499, arXiv:cond-mat/0207682.}
\nref\BMJ{M. Bousquet-M\'elou and A. Jehanne,
{\it Polynomial equations with one catalytic variable, algebraic
series and map enumeration}, J. Combin.\ Theory
Ser.\ {\bf B 96} (2006) 623-672.}
\nref\HANKEL{J. Bouttier and E. Guitter, {\it Planar maps and continued 
fractions}, Commun. Math. Phys. {\bf 309(3)} (2012) 623-662, arXiv:1007.0419.}
\nref\TUTEQ{W.T. Tutte, {\it On the enumeration of planar maps}, Bull.\
Amer.\ Math.\ Soc.\ {\bf 74} (1968) 64-74.}

\newsec{Introduction}

The study of planar maps, i.e.\ proper embeddings of graphs in the
two-dimensional sphere, has been the subject of an intense activity in
combinatorics since the seminal papers of Tutte in the 60's
\TutteCPM. Recent developments \LGMBuz\ entered the realm of probability
theory, by the study of particular ensembles of maps drawn 
according to some prescribed probability distribution. Such ensembles arise
naturally in physics where maps serve as discrete models for various types
of fluctuating surfaces such as fluid membranes in soft condensed matter physics
or quantum space-times in the theory of two-dimensional quantum gravity. In these
contexts, maps have often been equipped with additional degrees of freedom such 
as spins or particles, giving rise to nice universal critical behaviours
in the limit of large maps \DGZ. A particular important class of such models are 
the so-called $O(n)$ loop models, where maps carry self- and mutually-avoiding loops, 
each weighted by $n$ [\xref\DK-\xref\BE]. New universality classes are reached in 
these $O(n)$ loop models when the loops become large and modify the statistics 
of the underlying map. While the critical exponents have been determined
exactly and corroborate the celebrated KPZ relations \KPZ, which relate them to critical exponents of the $O(n)$ model on a fixed regular lattice, little is known on the
random geometry of these models.

This has to be contrasted with the case of planar maps without additional
degrees of freedom, whose geometry is now well understood. In
particular, when maps are endowed with the graph distance, thus are
viewed as metric spaces, their generic scaling limit (in the
Gromov-Hausdorff sense) is described by a unique remarkable
probabilistic object: the Brownian map
[\xref\LGMBuz,\xref\MierUniq,\xref\LGUniq]. Physically, it corresponds to the
universality class of so-called pure gravity. One may escape from this
universality class by considering maps with large faces (or ``holes'')
\LGM. More precisely, one works with the so-called Boltzmann ensemble
of random maps, where each face receives a weight depending on its
degree. If the face degrees are bounded (i.e.\ the weight for faces of
degree $k$ is set to zero for $k$ large enough -- which is a usual
assumption), the scaling limit is the Brownian map. Non-generic
scaling limits may only be obtained by allowing for unbounded degrees,
and furthermore fine-tuning the weights in such a way that the degree
distribution of a typical face has a heavy tail characterized by an
exponent $\alpha \in ]1,2[$. Under this assumption, the scaling limit
is a so-called ``stable map'' of Hausdorff dimension $2\alpha \in
]2,4[$ \LGM, different from the Brownian map of dimension 4. At first,
the fine-tuning of the weights might seem slightly unnatural, but it
was proposed that it occurs spontaneously in the context of critical
$O(n)$ loop models on random maps: considering a sample configuration
of such a model, if one erases all the outermost loops and their
contents, the resulting map (called {\it gasket}) is a map drawn
according to a Boltzmann ensemble with non-generic scaling limit.

This mechanism was investigated in a recent paper \BBG, for a
particular class of $O(n)$ loop models on tetravalent maps (dual to
quadrangulations). Notably, $n$ was shown to be generically
related to $\alpha$ via $n = 2 \sin \pi \alpha$ (where $n$ should vary
between $0$ and $2$ in order for the loop model to have critical
points, the two corresponding possible values of $\alpha$ being
associated respectively with the dense and dilute universality classes).
A full explicit solution of the model was 
obtained for a specific model, the so-called {\it rigid
loop} model where loops are forced to go straight within each visited
square. The purpose of the present paper is to extend the study of
\BBG\ to the largest possible class of $O(n)$ loop models amenable to
a full solution by the same techniques. As we shall discover, this
class corresponds to models with loops visiting triangles only, and
with an additional {\it bending energy} term which controls the
rigidity of the loops. These models interpolate, upon varying the
bending energy, between the standard $O(n)$ loop model on
triangulations [\xref\DK-\xref\BE] and the rigid loop model of
\BBG\ on quadrangulations (with squares formed of two triangles). 
Note that here, we define our model only on planar maps. Extending and
solving it for maps of arbitrary topology would present no difficulty,
following the lines of \GBThese, although we do not address it here. 
Indeed, once the planar case is solved, the answer for other topologies 
should be given by a topological recursion formula.

The paper is organized as follows: in Section 2, we recall the nested loop
approach initiated in \BBG\ as a general strategy that allows to reformulate 
any $O(n)$ loop model as a model of maps with controlled face degrees. In practice,
the nested loop approach consists of a gasket decomposition (Section 2.1) obtained by cutting 
the maps along the superimposed loops, resulting in a bijective coding of the $O(n)$ loop 
configurations. This coding translates into a {\it fixed point condition} 
for the degree-dependent weights in
the equivalent problem of maps with controlled face degrees (Section 2.2). 
This in turn translates into a {\it functional equation} for the so-called {\it resolvent}
of the $O(n)$ loop model, which is the generating function for loop configurations
on maps with a boundary of controlled length (Section 2.3).
Here, in contrast with \BBG, we use a shortcut to directly write down the functional equation from the fixed point condition.
Explicit examples of functional equations are given in Section 2.4 for a number
specific models that have been considered before. In all generality, the functional
equation involves a bivariate ring generating function which is the (grand canonical)
generating function for the immediate surroundings of the loops. Section 3 is devoted to finding
the largest possible class of models for which the functional equation may
be solved by a straightforward generalization of the techniques presented in \BBG\ 
and originally developed by Kostov \Kost, Kostov and Staudacher \KostSta, Eynard and Zinn-Justin \EZJ\ and Eynard and 
Kristjansen [\xref\EK,\xref\EKmore], and reformulated later in terms of algebraic geometry \GBThese. These models correspond to having a bivariate ring generating
function with a single pole, as explained in Section 3.1, in which case the functional 
equation simplifies drastically and involves crucially some (general) decreasing homographic involution. 
Interestingly enough, all these models may be realized by considering a model of loops
visiting triangles only and weighted by some particular {\it bending energy}, as explained 
in Section 3.2. The details of the solving technique are finally recalled in Section 3.3. 
This strategy is applied explicitly in Section 4 to study the phase diagram of the $O(n)$ 
loop model with bending energy defined on triangulations, i.e. when the unvisited faces
are themselves triangles. An explicit derivation of the non-generic critical line,
corresponding to the location in the parameters of the model of its dense and dilute 
phases, is given in Section 4.1 while the whole phase diagram is discussed in 
Section 4.2. The limit $n\to 0$ of the model is discussed in Section 4.3
as a number a combinatorial simplifications arise in this case. 
Section 5 is devoted to the solution by similar techniques of the so-called
{\it twisting loop} model on quadrangulations which is yet another particular case of the model
considered in \BBG\ where, in contrast with the rigid loop model, loops
are forced to make a turn inside each visited square. 
Our derivation of the functional equation for the resolvent of the $O(n)$ model
relies on the analytic properties of the similar resolvent in a model of maps
with controlled face degrees. We review these properties in Section 6. We
present in Section 6.1 a new combinatorial proof of the so-called {\it one-cut lemma}
and rederive in Section 6.2 the functional equation satisfied by the resolvent.
Section 6.3 is devoted to the proof of a technical result while the case of complex-valued face weights is briefly discussed in Section 6.4.
We gather our conclusions in Section 7.

\newsec{The nested loop approach}

\subsec{The gasket decomposition}

In this paper, we consider loop configurations on planar maps defined as 
follows: given a map, a {\it loop} is an undirected simple closed path on the 
dual map (i.e. it visits edges and vertices of the dual map, hence visits faces
and crosses edges of the original map). A {\it loop configuration} is a set of 
disjoint loops. Choosing a loop configuration amounts to choosing a set of 
crossed edges in such a way that each face is incident to exactly $0$ or $2$ 
crossed edges (with multiplicities, to account for self-folded faces). We call {\it map with a boundary of length} $p$ ($p\geq 1$) a rooted planar map
where the {\it external} face (i.e. the face on the right of the root edge) has
degree $p$. By convention, when considering loop configurations 
on maps with a boundary, we demand that the external face is not visited
by a loop. The external face allows to distinguish the exterior and
interior of a loop. In particular, we may associate to each loop
its {\it outer} (resp. {\it inner}) {\it contour} consisting of all the 
exterior (resp. interior) edges that are incident to a face visited by the 
loop. Each contour is a closed, but non-necessarily simple, path on the map.

\fig{A schematic picture of the gasket decomposition. Cutting the map along the outer 
contour of each outermost loop and removing the interior of these outer contours yields
the gasket, with a new type of faces: the holes. Upon cutting along the inner
contours of the same loops, the former content of each hole may itself be divided
into a ring and an internal map. All these components are rooted by some 
canonical (yet irrelevant) procedure. We may for instance consider the leftmost 
loop-avoiding shortest path from the root edge of the map (i.e. the root of the 
external face) to the outer contour of any outermost loop. This selects an edge 
on this contour which can be used as root for the corresponding ring. By then 
transferring the root from the outer to the inner contour by some local rule, 
it gives a root for the corresponding internal map. 
}{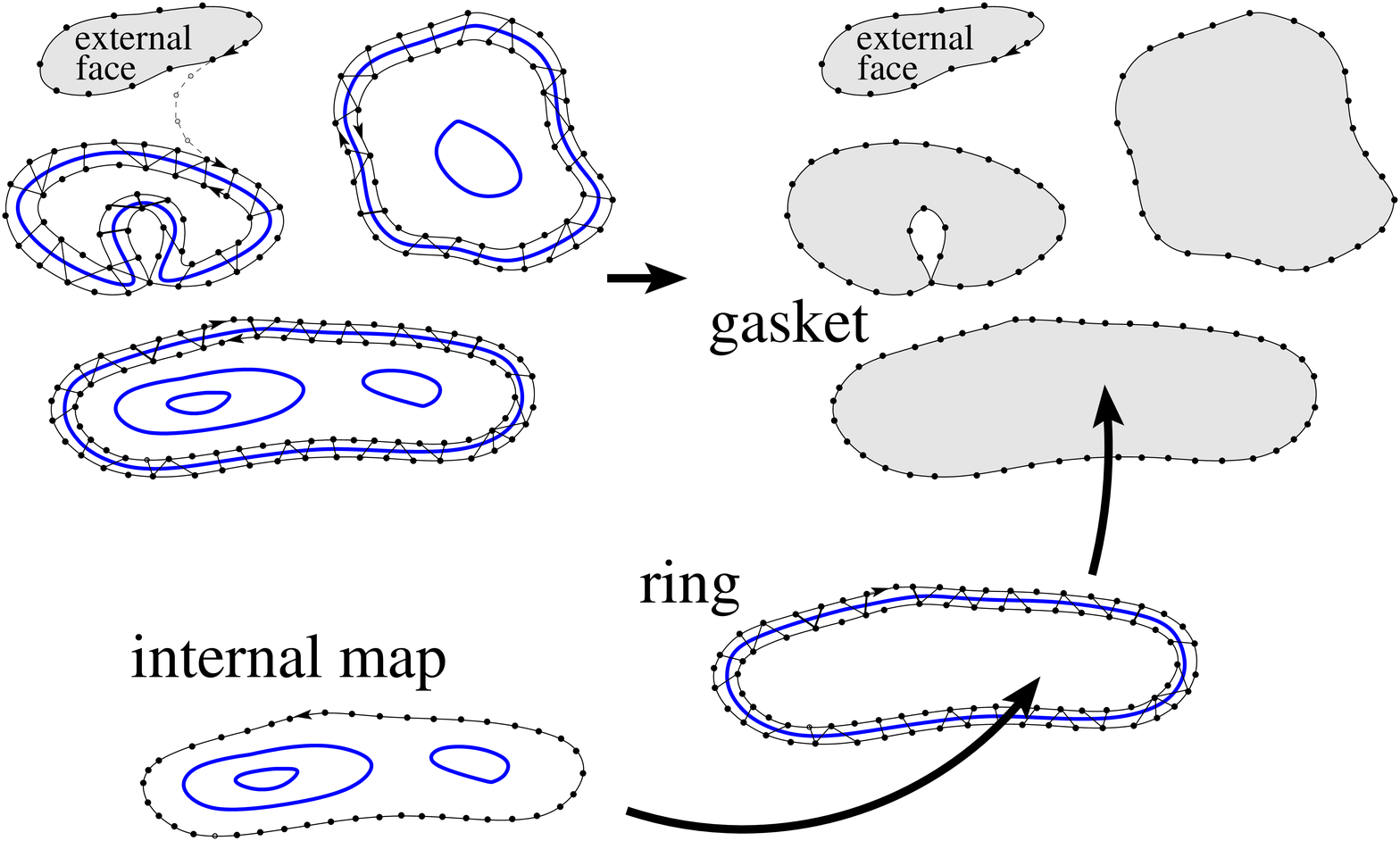}{13.cm}
\figlabel\gasketdecomp

Following \BBG, let us define the {\it gasket decomposition} of a map
with a boundary of length $p$ endowed with a loop configuration. It
consists in cutting the map along the outer and inner contours of each
{\it outermost} loop. This splits the map into several
connected components, which are as follows.
\item{-} The {\it gasket} is the map spanned by the 
edges which were exterior to all the loops.
\item{-} Each outermost loop yields two connected components:
\itemitem{-} its {\it ring} is formed by the faces visited by the loop, 
\itemitem{-} its {\it internal map} is the map spanned by the edges 
which were interior to the loop. This map is endowed with
a loop configuration consisting of all the loops which were originally 
interior to the outermost loop at hand.
\par\noindent All these components are rooted canonically, as displayed
in Fig.~\gasketdecomp. Thus, each internal map may be viewed as a map with a boundary
of the same nature as the original map.

By construction, the gasket is itself a map with a boundary of length $p$ 
(note that the gasket contains all the edges incident to the external face),
without loops, and whose inner faces are of two types: {\it regular} faces 
corresponding to faces of the original map that were exterior to all the 
loops, and {\it holes} delimited by the outer contours of the former 
outermost loops. Note that an outermost loop having outer and inner
contours of respective lengths $k$ and $k'$ gives rise to a hole
of degree $k$ in the gasket and to an internal map with a boundary of length
$k'$. Furthermore, both the gasket and internal map may contain
separating vertices. In particular, a multiple point along the outer
(resp. inner) contour yields a separating vertex incident to a hole in
the gasket (resp. to the external face of the internal map). By contrast, 
we decide that multiple points on the contours are split into distinct
vertices in the ring, so that both sides of the ring are simple paths
of lengths $k$ and $k'$.

To summarize, the gasket decomposition provides a bijection between,
on the one hand, maps with a boundary of length $p$ endowed with a loop
configuration, and on the other hand, maps with a boundary of length
$p$ with regular faces and holes, where each hole of degree $k$ ($k\geq 1$)
is endowed with (i) a ring with sides of lengths $k$ and $k'$ for 
some $k'\geq 0$ and (ii) a map with a boundary of length $k'$ endowed
with a loop configuration (by convention there is a unique map 
with a boundary of length $0$, the vertex-map, with no loops).

\subsec{The fixed-point condition}

We may use the above gasket decomposition to relate the partition function
of various loop models on random maps to the multivariate generating function 
for maps with prescribed face degrees. More precisely, consider an $O(n)$
loop model on maps with a boundary where the Boltzmann weight of a configuration
(i.e. a map endowed with a loop configuration) is the product of:
\item{-} for each $k \geq 1$, a weight $g_k^{(0)}$ per inner face of degree $k$ not visited by a loop,  
\item{-} a weight per loop that {\it only depends on the structure of its ring}, that we write in the form
$n\, w_r$ where $n$ is the loop fugacity and $w_r$ is, at this stage, 
an arbitrary function of the ring $r$.
In all the cases discussed in this paper, $w_r$ will be given as a product of
some local weights, leaving $n$ as the only non-local weight in the model.
\par\noindent
Let us denote by $F_p$ the partition function for such
an $O(n)$ loop model on maps with a boundary of length $p$,
which depends on $n$, the $g_k^{(0)}$'s and the $w_r$'s
(by convention, we set $F_0=1$). It follows from the gasket decomposition that 
$F_p$ is obtained by specializing a
multivariate generating function for maps without loops but with controlled face
degrees.
More precisely, we denote by ${\cal F}_p(g_1,g_2,\ldots)$
the multivariate generating function for maps with a boundary of length $p$ with
weight $g_k$ per inner face of degree $k$ (by convention ${\cal F}_0=1$),
and we introduce the {\it ring generating function}
\eqn\Adef{A_{k,k'} = \sum_{r \in {\cal R}(k,k')} w_r}
where ${\cal R}(k,k')$ is the set of rings with outer length $k$ and inner length $k'$,
rooted on the outer contour.
Then, the gasket decomposition translates into the relation
\eqn\FpFp{F_p={\cal F}_p(g_1, g_2,\ldots)}
at values of the weights $g_k$ satisfying the {\it fixed-point condition}
\eqn\fixedpoint{g_k=g_k^{(0)} + n \sum_{k'\geq 0} A_{k,k'} 
{\cal F}_{k'}(g_1,g_2,\ldots) \quad \hbox{for all} \ k\geq 1\ .}

Let us now give a few examples of ring generating functions. 
If we assume that the loops only visit triangles, and that the weight for a ring $r$ made of $m$ triangles is $w_r = h^m$, then we recover the $O(n)$ loop model studied in [\xref\Kost-\xref\EKmore]. We have
\eqn\Akktri{A_{k,k'}^{{\rm tri}}={k+k'-1 \choose k'} h^{k+k'} \ ,}
as seen by the following simple counting argument: a ring with
outer length $k$ and inner length $k'$ is a sequence of
$k+k'$ triangles, $k$ of which are facing outwards, that we may
read starting from the distinguished outward-facing triangle carrying
the root of the ring.

Consider now the $O(n)$ loop model studied in \BBG. Here, the loops only visit squares, and there are two types of visited squares: those such that the loop crosses opposite sides (type b) and those such that the loop crosses adjacent sides (type c). The weight for a ring $r$ made of $m$ squares of type b and $m'$ squares of type c is $w_r=h_1^m h_2^{m'}$, and the ring generating function is given by 
[\xref\BBG, eq.~(2.4)]
\eqn\Akkquad{A_{k,k'}^{\rm quad}=\sum_{j=0}^{\min(k,k')} {2k\over k+k'}{(k+k')!\over (2j)!(k-j)!(k'-j)!} h_1^{2j}
h_2^{k+k'-2j}\ .}
In particular, for $h_2=0$ we recover the so-called {\it rigid loop} model for which
\eqn\Akkrigid{A_{k,k'}^{{\rm rigid}}= \delta_{k,k'} h_1^k}
as seen directly by noting that the unique ring of outer length $k$ is made of $k$ squares and has
inner length $k'=k$.
Alternatively, let us consider the opposite case $h_1=0$ that we call the {\it twisting loop} model. Clearly, the associated ring generating function $A_{k,k'}^{\rm twist}$ vanishes whenever $k$ or $k'$ is odd, and
\eqn\Akktwist{A_{2k,2k'}^{\rm twist}= 2 {k+k'-1 \choose k'} h_2^{k+k'} \ ,}
as seen directly by a counting argument similar to that for \Akktri\ (there are $2$ possible choices for the position of the root on its incident outward-facing square, thus the extra factor $2$).

In general, we may encode the $A_{k,k'}$'s into the grand generating function
\eqn\bivA{A(x,y)=\sum_{k\geq 1} \sum_{k'\geq 0} A_{k,k'} x^k y^{k'}\ .}
The apparent asymmetry between $k$ and $k'$ in the above expressions
is related to the fact that the ring has a distinguished edge
on the outer contour. Assuming that the ring weight does not depend on the position
of the distinguished edge, we may write
\eqn\Aform{A(x,y)=x {\partial \over \partial x} \log H(x,y)}
where $\log H(x,y)$ is now the generating function for unrooted rings 
(involving symmetry factors).  If the ring weight is invariant when 
exchanging the exterior and the interior, then
$H(x,y)$ is symmetric in $x$ and $y$.
In all the cases discussed in this paper, $H(x,y)$ thus $A(x,y)$ will
be simple rationals function of $x$ and $y$. For instance, in the
triangular, quadrangular, rigid and twisting cases discussed above, we have
respectively
\eqn\HAex{\eqalign{H^{\rm tri}(x,y)={1 \over 1 - h (x+y)}, &\qquad
A^{\rm tri}(x,y)={h x \over 1 - h (x+y)}, \cr
H^{\rm quad}(x,y)={1 \over 1 - h_1 x y - h_2 (x^2+y^2)}, &\qquad
A^{\rm quad}(x,y)={h_1 x y + 2 h_2 x^2 \over 1 - h_1 x y - h_2 (x^2+y^2)}, \cr
H^{\rm rigid}(x,y)={1 \over 1 - h_1 x y}, &\qquad
A^{\rm rigid}(x,y)={h_1 x y \over 1 - h_1 x y}, \cr
H^{\rm twist}(x,y)={1 \over 1 - h_2 (x^2+y^2)}, &\qquad
A^{\rm twist}(x,y)={2 h_2 x^2 \over 1 - h_2 (x^2+y^2)}.}}
More generally, if the ring weight is simply given by a product of local face 
weights ({\it lfw}), say $h_{\ell,\ell'}$ per face incident to $\ell$ external 
edges and $\ell'$ internal edges, then $H(x,y)$ corresponds to 
the generating function for unrooted rings with a distinguished face.
Such rings are in correspondence with sequences of faces, so that
\eqn\HAgen{H^{\rm lfw}(x,y) = {1 \over 1 - \sum_{\ell,\ell' \geq 0} h_{\ell,\ell'} x^\ell y^{\ell'}}, \qquad
A^{\rm lfw}(x,y) = {\sum_{\ell,\ell' \geq 0} \ell h_{\ell,\ell'} x^\ell y^{\ell'} \over 1 - \sum_{\ell,\ell' \geq 0} h_{\ell,\ell'} x^\ell y^{\ell'}}.}
which are rational if the family $(h_{\ell,\ell'})_{\ell,\ell' \geq 0}$ has finite support. The symmetric case corresponds to
having $h_{\ell,\ell'}=h_{\ell',\ell}$ for all $\ell,\ell' \geq 0$.

\subsec{The functional equation for $O(n)$ loop models}

The fixed point condition \fixedpoint, which is an equation  for the infinite sequence $(g_k)_{k\geq 1}$, may be rephrased
as a functional equation for the {\it resolvent} of the model,
defined as
\eqn\resolv{W(x)=\sum_{p\geq 0} {F_p\over x^{p+1}}\ .}
Recall that $F_p$ is a specialization of the generating function ${\cal F}_p(g_1,g_2,\dots)$
for maps with controlled face degrees. As such, the resolvent $W(x)$ satisfies a number
of known analytic properties, which we review in Section 6. 
In particular, when the model is well-defined, that is to say the parameters of the
model ($n$, the $g_k^{(0)}$'s and the $w_r$'s) are such that the $F_p$'s are finite, 
$W$ may be analytically continued into a function that is holomorphic on the complex plane 
except on a cut where it has a finite 
discontinuity (see Section 6.1). When all weights are real non-negative, this cut is a real interval
$[\gamma_-,\gamma_+]$ with $\gamma_+\geq |\gamma_-|$. 
The discontinuity of $W$ on its cut is the so-called {\it spectral density}
\eqn\Wtorho{\rho(x)={W(x-{\rm i}0)-W(x+{\rm i}0)\over 2{\rm i}\pi}}
which vanishes at $x=\gamma_\pm$.
Moreover, as discussed in Section 6.2, we have the fundamental relation 
\eqn\cut{W(x+{\rm i}0)+W(x-{\rm i}0)= V'(x)\ , \quad x \in ]\gamma_-,\gamma_+[\ ,}
where we introduce the shorthand notation
\eqn\Vprime{V'(x)= x -\sum_{k\geq 1} g_k x^{k-1}\ .}
This fundamental relation holds for the resolvent of general maps with
controlled face degrees, i.e.\ does not require that the weight
sequence $(g_k)_{k\geq 1}$ satisfies the fixed-point condition
\fixedpoint. This latter condition \fixedpoint\ allows to rewrite $V'(x)$ as  
\eqn\potB{\eqalign{V'(x) & = x - \sum_{k \geq 1} g_k^{(0)}\,x^{k - 1} - n\sum_{k \geq 1}\sum_{k' \geq 0} A_{k,k'}\,F_{k'}\,x^{k - 1} \cr
& = V_0'(x) - {n \over 2{\rm i}\pi x}\oint_{{\cal C}}\,A(x,y)\,W(y)\,dy}}
where $A(x,y)$ is defined as in \bivA, ${\cal C}$ is a contour surrounding 
$[\gamma_-,\gamma_+]$ and
\eqn\Vprimezero{V_0'(x)=x-\sum_{k\geq 1}  g_k^{(0)} x^{k-1}\ .}
Note that the contour ${\cal C}$ must be included in the domain of analyticity 
of $y\mapsto A(x,y)$ for all $x \in ]\gamma_-,\gamma_+[$.
We end up with the linear functional equation
\eqn\cutbis{W(x+{\rm i}0)+W(x-{\rm i}0)
= V_0'(x) - {n \over 2 {\rm
i} \pi x} \oint_{\cal C} A(x,y) W(y) dy \ , \quad x \in ]\gamma_-,\gamma_+[\ .}
A new feature of this equation is that $W$ now appears also on the rhs.

\subsec{Some examples of functional relations} 

When $A(x,y)$ is a rational function, we may evaluate the integral in \potB\ 
using the residue theorem. For instance, in the triangular case,
$A^{\rm tri}(x,y) W(y)$ has by \HAex\ one simple pole at $y=h^{-1}-x$,
hence we recover the functional equation [\xref\Kost-\xref\EKmore] 
\eqn\cuttri{W^{\rm tri}(x+{\rm i}0)+W^{\rm tri}(x-{\rm i}0) = V_0'(x)- n\, W^{\rm tri}(h^{-1} - x).}
Similarly in the rigid case, $A^{\rm rigid}(x,y) W(y)$ has one simple pole at $y=1/(h_1 x)$ and
a residue $1$ at infinity (since $W(y) \sim 1/y$), hence we recover the functional equation \BBG 
\eqn\cutrigid{W^{\rm rigid}(x+{\rm i}0)+W^{\rm rigid}(x-{\rm i}0) = V_0'(x)  + {n \over x} - {n \over h_1 x^2} W^{\rm
rigid}\left( {1 \over h_1 x} \right) \ .}
In the generic quadrangular case ($h_2>0$), $A^{\rm quad}(x,y)\,W(y)$ has two simple poles at the roots
\eqn\roof{y_\pm(x) = {-h_1x \pm \sqrt{(h_1^2 - 4h_2^2)x^2 + 4h_2} \over 2h_2}}
and the functional equation reads
\eqn\cutquad{\eqalign{& W^{\rm quad}(x+{\rm i}0)+W^{\rm quad}(x-{\rm i}0) = \cr & \qquad
V_0'(x) + n\, y_+'(x)\, W^{\rm quad}(y_+(x)) + n\, y_-'(x)\, W^{\rm quad}(y_-(x))\ .}}

Let us now discuss the case where the gasket is a bipartite map, i.e.\ when all its faces have even degree.
Such a situation occurs whenever $g_{2k+1}^{(0)}$ and $A_{2k+1,2k'}$ vanish for all $k,k' \geq 0$, so
that $g_{2k+1}=0$ for all $k$. Then $F_p$ vanishes for all $p$ odd so that
$W(x)$ is an odd function of $x$ (in particular, $\gamma_-=-\gamma_+$). We may then
write $W(x) = x\,\tilde{W}(x^2)$
where $\tilde{W}(X)$ has a cut $[0,\Gamma]$ (with
$\Gamma=\gamma_{+}^2 = \gamma_-^2$) on which it satisfies
\eqn\cutbip{\tilde{W}(X+{\rm i}0)+\tilde{W}(X-{\rm i}0)= \tilde{V}'(X), \quad X \in [0,\Gamma[}
with $\tilde{V}'(X)=1-\sum_{k\geq 1} g_{2k} X^{k-1}$.
This equation is the bipartite counterpart of \cut. 
The fixed point condition \fixedpoint\ then yields the functional equation
\eqn\cutbisbip{\tilde{W}(X+{\rm i}0)+\tilde{W}(X-{\rm i}0)= \tilde{V}_0'(X) - {n \over 2 {\rm
i} \pi X} \oint \tilde{A}(X,Y) \tilde{W}(Y) dY}
where the integral is over a contour encircling the cut $[0,\Gamma]$ and
\eqn\Atildef{\tilde{V}_0'(X)=1-\sum_{k\geq 1} g_{2k}^{(0)}X^{k-1}\ , \qquad 
\tilde{A}(X,Y)=\sum_{k \geq 1} \sum_{k' \geq 0}
A_{2k,2k'} X^k Y^{k'}\ .}
In particular, in the twisting loop model, $\tilde{A}^{\rm
twist}(X,Y)=A^{\rm twist}(\sqrt{X},\sqrt{Y})$ has by \HAex\ a simple
pole at $Y=h_2^{-1}-X$ with residue $2$ so that the functional
equation reads
\eqn\cuttwist{\tilde{W}(X+{\rm i}0)+\tilde{W}(X-{\rm i}0)= \tilde{V}'_0(X) - 2 n\, \tilde{W}(h_2^{-1} - X)\
,}
which is very similar to \cuttri. 
The solution of the triangular case \cuttri\ was the topic of 
the original work of Kostov \Kost, and Eynard and Kristjansen [\xref\EK,\xref\EKmore]. 
With identical techniques, the solution for the twisting loop model on 
quadrangulations \cuttwist\ can be easily deduced (see Section 5). 
With a few modifications, the method was adapted to solve \cutrigid\ in \BBG.
In this article, we show how to solve via similar methods a more
general family of models (which contains those three examples as 
special cases). 

\newsec{The $O(n)$ loop model on cubic maps with bending energy: equations}

\subsec{Ring generating functions with one pole}

A peculiar feature of the triangular and rigid loop models 
is that both the ring generating functions $A^{\rm tri}(x,y)$ and
$A^{\rm rigid}(x,y)$, as defined in \HAex, are rational functions with one pole, which leads
to a simplification of the functional equation \cutbis\ into
respectively \cuttri\ and \cutrigid. In this section, we put those two
models under the same roof by considering the general ``one-pole
case'' (note that the bipartite ring generating function
$\tilde{A}^{\rm twist}(X,Y)$ is also a one-pole rational function but,
for simplicity, we postpone the discussion of the twisting loop model
to Section 5).

More precisely, let us first assume that the ring weight is symmetric
and does not depend on the position of the distinguished edge, so that
$A(x,y)$ is given by \Aform, with $H(x,y)$ a symmetric function. Let
us moreover assume that $H(x,y)$ be a one-pole rational function
of the form 
\eqn\Hsimpol{H(x,y) = {h(y) \over x - s(y)},}
so that $A(x,y)$ is also a one-pole rational function in $x$.
Note that to fulfill this latter condition, it is important that
the numerator does not depend on $x$, since any zero of the numerator
would give
rise to an extra pole in $A(x,y)$. Demanding that $H(x,y)$ be symmetric
in $x,y$ yields a functional equation for $h$ and $s$. It is readily
solved (e.g. write the equation for $y=0$ and solve for $h(x)$, then
substitute into the equation for $y=1$) and implies that $s(x)$ takes
the form 
\eqn\Shomo{s(x)= {\alpha-\beta x \over \beta -\delta x}} 
for some real parameters $\alpha$, $\beta$ and $\delta$, while
$h(x)\propto 1/(\alpha-\beta x)$. Note the remarkable property
$s(s(x))=x$. In other words, we recognize in \Shomo\ the general
expression for a {\it homographic involution}. In particular,
this involution reduces to 
$s(x)=h^{-1}-x$ for the triangular model and to $s(x)=1/(h_1 x)$ for the
rigid model. By \Aform, we have
\eqn\formAxy{A(x,y)={x\over s(y)-x}}
hence the ring generating
function $A_{k,k'}$ satisfies the {\it exponentiation property}
\eqn\facA{\sum_{k'\geq 0} A_{k,k'}y^{k'}= s(y)^{-k}, \qquad k \geq 1.}
For positive ring weights, $s(y)$ is then necessarily a decreasing
positive function of the real positive variable $y$. This imposes $s(0)>0$
together with the extra condition
$\beta^2-\alpha \delta>0$. Hence, the involution has two real fixed
points at $(\beta \pm \sqrt{\beta^2-\alpha \delta})/\delta$.  We will
see in the next subsection that all the decreasing homographic
involutions are indeed obtained in a simple model of loops with
bending energy. In Appendix A, we mention the cases where the homographic 
involution is increasing, although they cannot be reached with non-negative 
local weights.

By the form \Shomo\ for $s(x)$, one may easily check that $A(x,y)$,
as given by \formAxy, may be written in the alternative form
\eqn\formAxyalt{A(x,y)={x s'(x)\over y-s(x)}+ {x s''(x)\over 2 s'(x)}\ ,}
which displays its single pole in the variable $y$ at $y=s^{-1}(x)=s(x)$.
We may now evaluate the contour integral in \cutbis\ via the residue
theorem.  Recall that, in order for the model to be well-defined, the
cut $[\gamma_-,\gamma_+]$ of $W$ has to be contained, for all $x$ on
the cut, within the disk of convergence of $y \mapsto A(x,y)$ which
here has radius $|s(x)|$. This shows that the interior of the cut may
not overlap with its image under $s$ (but the endpoints of the cut may
be fixed points of $s$).
The integrand $A(x,y) W(y)$ has a pole at $y=s(x)$, with residue equal
to $-x s'(x) W(s(x))$. Since $W(y)\sim 1/y$ for $y\to \infty$, the
residue at infinity is equal to $\lim_{y\to \infty} A(x,y)=x s''(x)/(2
s'(x))$. We deduce that
\eqn\valint{{1 \over 2 {\rm i} \pi} \oint A(x,y) W(y) dy  
= -x s'(x) W(s(x)) +{x s''(x)\over 2 s'(x)}}
which, inserted in \cutbis, yields the functional equation
\eqn\funceq{W(x+{\rm i}0) +W(x-{\rm i}0)- n s'(x)W(s(x))= V'_0(x)
- {n s''(x)\over 2 s'(x)}\ , \quad x \in ]\gamma_-,\gamma_+[}
This equation generalizes eqs.~\cuttri\ and \cutrigid, which
are recovered respectively for $s(x)=h^{-1}-x$ and $s(x)=1/(h_1 x)$, 
as expected.

\subsec{Realization in a model of loops with bending energy}

In this Section, we exhibit a model whose ring generating function
reproduces precisely the most general symmetric one-pole rational function
of the previous subsection,
hence whose resolvent is the solution of the general functional equation
\funceq.  We consider again an $O(n)$ loop model where the
loops visit only triangles, with a weight $h$ per visited triangle,
but we also introduce a {\it bending energy} for the loops in the
following way: each triangle visited by a loop has exactly one edge
which is not crossed by the loop. We say that the triangle faces
outwards (resp. inwards) if this edge is on the same side of the loop
as the external face (resp. on the opposite side). Now for each {\it
pair} of successive triangles along a loop, we attach a weight $a$ if
the triangles are of the same nature (either both facing outwards or
both facing inwards) and a weight $1$ otherwise. We may interpret
$-\log a$ as a bending energy for the loops as this new weight favors
loop turns for $a>1$ and penalizes them for $0<a<1$.  Such a bending
energy was previously introduced in \Lorentz\ in the different context
of Lorentzian triangulations. Note that our model precisely interpolates
between the regular $O(n)$ loop model
on triangles, recovered for $a=1$, and the rigid $O(n)$ loop model on
squares, recovered for $a=0$ and $h_1=h^2$. Indeed, at $a=0$, outward-
and inward-facing triangles necessarily alternate along the loop
and, upon concatenating them by pairs, we may simply replace them by
squares (with weight $h^2$) having opposite sides crossed by the loop.

Let us now compute the ring grand generating function $A(x,y)$. 
Rings whose inner contour has zero length are
non-empty sequences of outward-facing triangle only, each triangle
receiving a weight $ahx$. These rings therefore have a grand generating
function equal to ${ahx \over 1-ahx}$. Rings with inner contour of non-zero length
may be viewed, before choosing their root edge, as cyclic sequences of 
composite objects consisting of (i) a non-empty sequence of outward-facing 
triangles followed by (ii) a non-empty sequence of inward-facing triangles. 
Each composite object therefore yields a grand generating function 
${hx \over 1-ahx}\cdot {hy \over 1-ahy}$ (with no weight $a$ for the first triangle
of each sequence since it follows a triangle of the opposite type). We deduce
eventually
\eqn\Axybending{A(x,y)= {a h x\over 1-a h x}+ 
x{\partial \over \partial x}\left(
-\log\left(1- {hx \over 1- a h x}{hy\over 1-a hy}\right)\right)}
where, in the second term, the $\log(\cdot)$ accounts for the cyclic 
sequence of composite objects while the $x\partial_x$ accounts for the 
rooting of the ring. We observe that $A(x,y)$ is indeed of the form
\Aform\ with
\eqn\Hxybending{H(x,y)= {1 \over 1 - a h (x+y) - (1-a^2) h^2 x y}}
which is a symmetric one-pole rational function (observe the
similarity with Eq.~(2.3) of \Lorentz).
We are therefore
precisely in the situation discussed in Section 3.1, hence
we may rewrite $A(x,y)$
in the forms \formAxy\ and \formAxyalt\ with $s(x)$ now given by 
\eqn\Sbending{s(x)={1-ahx\over ah+(1-a^2)h^2 x}}
which has two fixed points at $x={1 \over (a\pm 1)h}$. 
We therefore recover the general form \Shomo\ for an arbitrary
decreasing homographic involution such that $s(0) > 0$ (with ${1 \over s(0)} = {\beta \over \alpha}=ha$,
${\delta \over \alpha}=(a^2-1)h^2$, $\beta^2-\alpha\delta= h^2 \alpha^2>0$). In other words, our loop model with bending energy
realizes precisely the general ``one-pole'' case.
Note that $s(x)$ reduces to $h^{-1}-x$ for $a=1$ and to ${1 \over h^2 x}$
for $a=0$, as it should. 

\subsec{Solving strategy}

The technique developed by Eynard and Kristjansen \EK\ to solve the
functional equation \cuttri\ may be adapted to \funceq\ with few
modifications, as we will now sketch. Briefly, we may see that,
for fixed given $\gamma_\pm$, the
functional equation admits a unique solution $W$ analytic outside
the cut $[\gamma_-,\gamma_+]$ and such that $W(x) \sim 1/x$ for $x \to
\infty$. Then, the unknowns $\gamma_\pm$ are determined {\it a posteriori} by the condition
$\rho(\gamma_\pm)=0$, where $\rho$ is the spectral density \Wtorho. 

We will restrict ourselves to the case where $V_0'$ is a polynomial,
i.e. we assume that the degrees of the faces not
visited by loops are also bounded. 
In that case, $V_0'$ is defined on the whole complex plane, and we easily find a particular solution of \funceq\ with 
no discontinuity along the cut, namely 
\eqn\Wpart{W_{{\rm part}}(x) ={2 V'_0(x) + n s'(x) V'_0(s(x))\over 4-n^2} -
  {n s''(x)\over 2 (n+2) s'(x)}.} 
To check that this is indeed a particular solution, note that, $s(x)$
being an involution, we have $s'(s(x))=1/s'(x)$ and
$s''(s(x))=-s''(x)/s'(x)^3$. Since $V'_0$ is a polynomial, observe
that $W_{{\rm part}}$ has (multiple) poles at $\infty$ and at $s(\infty)$.
Any solution of \funceq\ is of the form 
\eqn\Wgeneral{W(x)=W_{{\rm part}}(x)+W_{{\rm hom}}(x)\ ,}
where $W_{{\rm hom}}$ is now a solution of the homogeneous 
functional equation
\eqn\homfunceq{W_{{\rm hom}}(x+{\rm i}0)+W_{{\rm hom}}(x-{\rm i}0)
-n\, s'(x) W_{{\rm hom}}(s(x))=0 \ ,
\quad x \in ]\gamma_-,\gamma_+[\ .}

Assuming that $s(\infty) \neq \infty$, viz. $\delta \neq 0$ in \Shomo\
and $a\neq 1$ in \Sbending,
the requirement $W(x) \sim 1/x$ for $x \to \infty$ yields the
condition
\eqn\condW{W_{{\rm hom}}(x)= - W_{{\rm part}}(x) + {1\over x} +
O\left({1\over x^2}\right)= -{2 V_0'(x)\over 4-n^2} +{2\over n+2}{1\over x} 
+O\left({1\over x^2}\right).}
Moreover, the analyticity of $W$ outside the cut, and the fact that $W$
is bounded in some neighborhood of the cut, require that the pole of
$W_{{\rm part}}$ at $s(\infty)$ is canceled, which may be rephrased as
\eqn\condWbis{\eqalign{s'(x) W_{{\rm hom}}(s(x))= - s'(x) W_{{\rm part}}(s(x)) 
+ O\left({1\over x^2}\right)& = -{n V_0'(x)\over 4-n^2} +{n\over n+2}{1\over x} 
+O\left({1\over x^2}\right)\cr & ={n\over 2}W_{{\rm hom}}(x)
+O\left({1\over x^2}\right) \cr} }
when $x\to \infty$. When $s(\infty)=\infty$, viz. $\delta = 0$ in
\Shomo\ or $a=1$ in \Sbending, the two 
conditions \condW\ and \condWbis\ are replaced by the
single condition
\eqn\condWspec{W_{{\rm hom}}(x)=- W_{{\rm part}}(x) + {1\over x} +
O\left({1\over x^2}\right)=-{2V_0'(x)+ ns'(x) V_0'(s(x))\over 4-n^2} 
+{1\over x} 
+O\left({1\over x^2}\right)\ .}

The general solution of \homfunceq\ may be expressed in terms of
elliptic functions: let us introduce the elliptic integral
\eqn\vint{v(x)=\int^{s(\gamma_-)}_x {d\xi \over \sqrt{\pm
(\xi-\gamma_+)(\xi-\gamma_-)(\xi-s(\gamma_+))(\xi-s(\gamma_-))}}\ .}
Since $s$ is (locally) decreasing and the cut $ [\gamma_-,\gamma_+]$ does not 
overlap with its image, the sign may be chosen in such a way that the polynomial
under the square root be positive over $[s(\gamma_+),s(\gamma_-)]$ 
(when $s(\gamma_+) > s(\gamma_-)$, this interval is to be understood as 
the union $]-\infty,s(\gamma_-)] \cup [s(\gamma_+),\infty[$). 
In particular, this implies that $T=v(s(\gamma_+))$ is real and positive. 
By the same reason, the polynomial under the square root is negative on 
$[s(\gamma_-),\gamma_-]$ (understood as $]-\infty, \gamma_-] \cup 
[s(\gamma_-),\infty[$ if $\gamma_-<s(\gamma_-$)) so that 
$v(\gamma_-)$ has two pure imaginary determinations $\pm {\rm i}T'$ ($T'>0$).
Finally, we may check that $v(\gamma_+)=T \pm {\rm i} T'$.

For the moment, we assume that $s(\gamma_\pm) \neq \gamma_\pm$ so that
$T<\infty$.  The function $x \mapsto v(x)$ has branch points of order
$2$ at $\gamma_\pm,s(\gamma_\pm)$ and maps the lower (resp.\ upper)
half-plane onto the rectangle $[0,T] + {\rm i} [0,T']$ (resp.\ $[0,T]
+ {\rm i} [-T',0]$). Its reciprocal function $v \mapsto x(v)$ may be
analytically continued to an even, doubly periodic function with
periods $2T$ and $2 {\rm i}T'$, with simple poles at 
$\pm v_\infty\ {\rm mod}\ 2T{\bf Z} + 2{\rm i}T'{\bf Z}$, where $v_\infty = \lim_{x \to \infty} v(x)$. Note the fundamental property
\eqn\vprop{x(v \pm {\rm i} T')=s(x(v))}
which follows from \vint: indeed, considering $v(s(x))$, the change of
variable $\xi \to s(\xi)$ leaves the integrand invariant (here it is
crucial that $s$ be a homographic involution) and sets the endpoints
of the integration path to $x$ and $\gamma_-$ respectively, so that
$v(s(x))=v(x) - v(\gamma_-)$.

For $v$ in the strip $]-\infty,\infty[ + {\rm i} ]-T',T'[$ (which is
mapped onto the complement of the cut $[\gamma_-,\gamma_+]$), let us
define
\eqn\whompara{\omega(v)=x'(v)\, W_{{\rm hom}}(x(v))\ .}
Since $W_{{\rm hom}}$ is meromorphic outside the cut
$[\gamma_-,\gamma_+]$, $\omega$ is meromorphic on the strip. Note that
$\omega$ is odd and $2T$-periodic. Furthermore, by the above
mapping property of $x$, we find that $\omega$ has well-defined values
on the boundary of the strip, namely
\eqn\omegalim{\omega(v \pm {\rm i} T') = x'(v \pm {\rm i} T')\,
W_{{\rm hom}}(x(v \pm {\rm i} T') \mp {\rm i} 0), \qquad 0 < v < T}
(the value of $\omega(v \pm {\rm i} T')$ for other real values of $v$ 
follows by oddness and $2T$-periodicity). By \homfunceq\ and
\vprop\ (and its derivative), we deduce that, for any real $v$,
\eqn\omegafunceq{\omega(v+{\rm i} T')+\omega(v-{\rm i} T')-n\, \omega(v)=0\ .}
This relation implies that $\omega$ may be analytically continued into
a meromorphic function defined on the whole complex plane, so that
\omegafunceq\ holds for all complex values of $v$. For simplicity, we assume now 
$n \neq \pm 2$ (for $n = -2,2$, the solution is also known, see for instance \KostSta\ 
but we will not discuss it here). As discussed in \BBG, 
any $2T$-periodic meromorphic solution of \omegafunceq\ can be expressed in terms 
of the ``fundamental'' solutions $\zeta_{\pm b}$, where $\pi b=\arccos (n/2)$ and
\eqn\theta{\zeta_b(v)= {\vartheta_1'(0|\tau) \over 2T\,\vartheta_1\left(-{b \over 2}|\tau\right)}\,{\vartheta_1\left({v \over 2T}  - {b \over 2}|\tau\right) \over \vartheta_1\left({v \over 2T}|\tau\right)},\qquad \tau = {{\rm i}T' \over T}}
(see the Appendix in \BBG\ for some properties of $\zeta_b(v)$). 
More precisely, any $2T$-periodic solution of \omegafunceq\
is a linear combination of derivatives of translates of $\zeta_{\pm b}$. 
Here, we are looking for an odd solution with
poles at $\pm v_\infty$ whose residues are fixed, when $s(\infty) \neq \infty$,
 by the condition
\eqn\condomega{\omega(v) =
x'(v)\left(- {2 V_0'(x(v))\over 4-n^2} +{2\over n+2}{1\over x(v)}\right)+
o\left({1\over v - v_\infty}\right)\qquad  {\rm for}\ v \to 
v_\infty}
inherited from \condW, as well as poles at $\pm v_\infty \pm {\rm i} T'$
with {\it the same residues multiplied by}  $n/2= \cos\pi b$ by virtue of \condWbis.
The general form for such a solution is 
\eqn\omegagensol{\omega(v) = \sum_{k \geq 0} a_k \left(
\zeta^{(k)}(v-v_\infty) - \zeta^{(k)}(-v-v_\infty) \right)}
where
\eqn\zetagen{\zeta(v)={\zeta_b(v)+\zeta_{-b}(v)\over 2}\ ,}
and where the coefficients $a_k$ are determined by \condomega.
Note that the particular combination $\zeta$ was chosen as it has residue 
$1$ at $0$ and residue $n/2$ at $\pm {\rm i} T'$ so that \condomega\ 
implies both \condW\ and \condWbis. 
In the case $s(\infty)=\infty$ (hence $v_\infty={\rm i} T'/2$), the condition \condomega\ 
has to be modified according to \condWspec.
We may observe that only a finite number of $a_k$'s are non-zero
when $V'_0$ is a polynomial. More precisely, when the maximal allowed face degree is $D$, 
then $a_k=0$ for $k>D$. 

After determining $\omega$ in this way, it remains
to satisfy the conditions $\rho(\gamma_\pm)=0$. By \omegalim, 
we may rewrite the spectral density \Wtorho\ in the
parametric form
\eqn\rhopara{\rho(x (v+ {\rm i} T')) = {\omega (v + {\rm i} T') -
\omega (v - {\rm i} T')\over 2 {\rm i} \pi\, x'(v+ {\rm i} T')}, 
\qquad 0 < v < T.}
Since $x$ has ramifications points of order 2 at ${\rm i} T'$ and $T+ {\rm
i} T'$, the conditions $\rho(\gamma_\pm)=0$ amount to $\omega({\rm i}
T')=\omega(T+{\rm i} T')=0$, which implicitly determine
$\gamma_\pm$.

We further deduce that $\rho$ (hence $W$) has generically a
square-root singularity at $\gamma_{\pm}$, corresponding to a
non-critical point in the terminology reviewed in Section 6.1. 
For $T<\infty$, a critical point may only be
obtained when $\omega''$ vanishes at either ${\rm i} T'$ or $T+ {\rm
i} T'$, so that $\rho$ now decays generically with an exponent $3/2$
at one endpoint of the cut: this corresponds to a generic critical
point, in the universality class of pure gravity. Multicritical points
can also be obtained, but only at the price of introducing negative
weights in the model. 

\fig{The (multivalued) change of variable from $x$ to $v$, as given by \vint\ 
in the particular case $\gamma_+=s(\gamma_+)$, up some normalization factor which
ensures that $v(\gamma_-)=\pm{\rm i} \pi$. The reciprocal function is
given by Eq.~(3.27). Here we assumed that $s(\infty)$ belongs to the cut
$]\gamma_-,\gamma_+[$ (solid line) so that the image (in the $x$-plane) 
of this cut by the involution $s$ is $]-\infty,s(\gamma_-)[\cup
]\gamma_+,\infty[$ (dashed line). The cut maps
onto the lines ${\rm Im}\,v=\pm \pi$ in the $v$-plane (solid lines), 
while its image maps onto the line ${\rm Im}\,v=0$ (dashed line).}{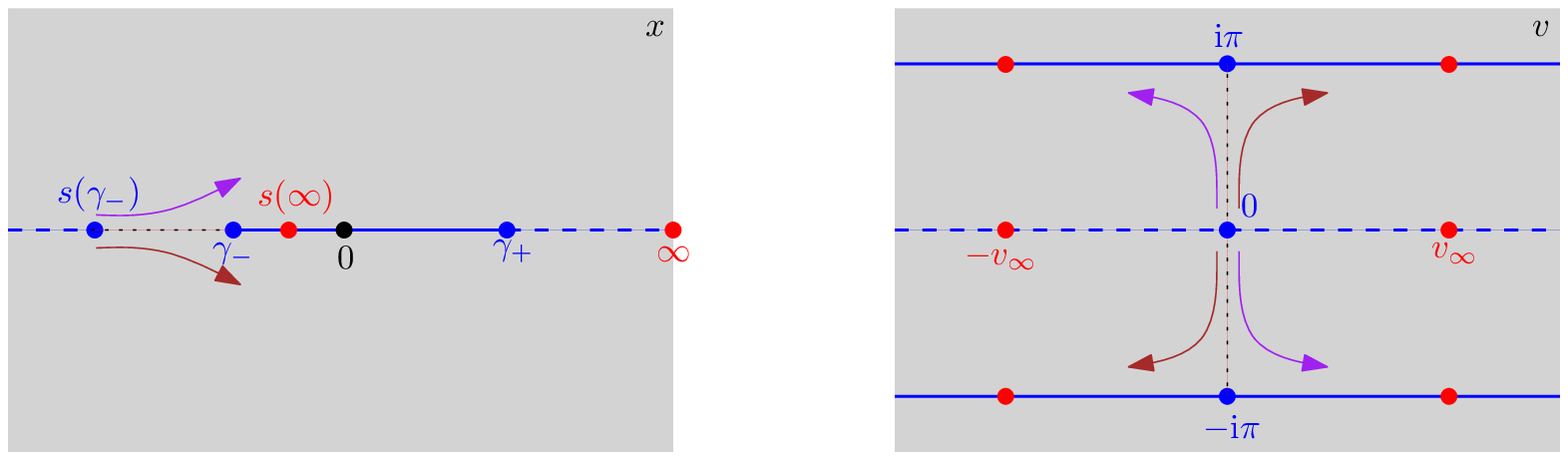}{14.cm}
\figlabel\trigparam

The only possible way to obtain a non-generic
critical point with non-negative weights is to have
$\gamma_+=s(\gamma_+)$, so that $T=\infty$.
The above strategy simplifies greatly in this
situation. Indeed, since one of the periods ($2T$) is infinite, the elliptic
(doubly periodic) parametrization degenerates into a (simply periodic)
trigonometric parametrization (see Fig.~\trigparam). It is then convenient to rescale $v$ by
a constant factor so that $v(\gamma_-)=\pm{\rm i} \pi$ (i.e. $T'=\pi$). 
The reciprocal function $v \mapsto x(v)$ then reads explicitly
\eqn\vtrigpara{x(v) = \left( \gamma_+ - s(\gamma_-) \right) 
  {\cosh v - 1 \over \cosh v - \cosh v_\infty} + s(\gamma_-)}
where $v_\infty$ is determined by the condition $x({\rm i} \pi)=\gamma_-$.
Furthermore, for $T=\infty$ and $0<n<2$, the fundamental solution $\zeta$ in \zetagen\ reads
\eqn\zetab{\zeta(v)=\cosh(b\, v)\coth v-\sinh(b\, v)}
with
\eqn\valb{\pi b = \arccos \left({n\over 2}\right)\,.}
Hence we may apply the above strategy on very explicit formulas.  The
only caution we must take concerns the condition $\rho(\gamma_+)=0$ which, since $x'(v)
\sim C e^{-v}$ for $v \to \infty$, now amounts to the condition
$\omega(v)=o(e^{-v})$. Noting that $\zeta$ has the asymptotic
expansion
\eqn\zetasymp{\zeta(v) = e^{-b v} + e^{-(2-b)v} + e^{-(2+b)v} + o(e^{-(2+b)v})}
with $0<b<1/2$, this yields a linear condition on the $a_k$'s appearing
in \omegagensol, that replaces the previous condition $\omega(T+{\rm
i}T')=0$.

Interestingly, this asymptotic expansion also yields the non-generic
critical exponents. Indeed, we generically have $\omega(v) \sim C'
e^{-(2-b)v}$ for $v \to \infty$ and, noting that $x(v)=\gamma_++\alpha
e^{-v}+o(e^{-v})$ with $\alpha>0$ (since $x(v)$ approaches $\gamma_+$
from the right when $v$ takes large real values), we deduce that
$W_{{\rm hom}}(x) \sim - \alpha^{b-2} C' (x-\gamma_+)^{1-b}$.
This is nothing but the leading singular term in the expansion of $W$
around $\gamma_+$, and the exponent $1-b$ is characteristic of the
universality class of the {\it dense $O(n)$ loop model on random maps}. By
transfer theorems, it amounts to the asymptotic behaviour
\eqn\Fkdense{F_k \sim - { \gamma_+^{1-b} C' \over \alpha^{2-b} \Gamma(b-1)}
{\gamma_+^k \over k^{2-b}}, \qquad k \to \infty.}
Since $0<b<1/2$, the positivity of $F_k$ (in a model with non-negative
weights) requires that $C' \geq 0$. When $C'=0$, we instead have
$\omega(v) \sim C'' e^{-(2+b)v}$ so that $W_{{\rm hom}}(x) \sim - \alpha^{-b-2} C''
(x-\gamma_+)^{1+b}$, and the exponent $1+b$ is characteristic of the
universality class of the {\it dilute $O(n)$ loop model on random
maps}. It corresponds to the asymptotic behaviour
\eqn\Fkdilute{F_k \sim - { \gamma_+^{1+b} C'' \over \alpha^{2+b} \Gamma(-(b+1))}
{\gamma_+^k \over k^{2+b}}, \qquad k \to \infty}
hence $C'' < 0$ in a model with non-negative weights. It is not
possible to have both $C'=0$ and $C''=0$ in a such a model, thus
\Fkdense\ and \Fkdilute\ are the only possible ``physical'' non-generic
critical behaviours.

\newsec{The $O(n)$ loop model on cubic maps with bending energy: phase diagram}

\subsec{The non-generic critical line}

In this section, we compute the non-generic critical line of the
$O(n)$ loop model with bending energy defined in Section 3.2.
As discussed in Section 3.3, non-generic criticality 
amounts to the condition $s(\gamma_+)=\gamma_+$.
Note that $\gamma_+$ is a continuous increasing function of the weights $g_k^{(0)}$ and $h$, with $\gamma_+=0$
when the weights vanish. Then, when $h$ is positive and $a$ non-negative, this condition occurs when 
$\gamma_+$ reaches the smallest (in modulus) fixed point of the involution, namely
\eqn\coales{\gamma_+={1\over (a+1) h}\ .}
Note that the other (largest in modulus) fixed point ${1 \over (a-1)h}$ of the 
involution does not necessarily coincide with the other endpoint 
$\gamma_-$ of the cut, but lies outside of both the cut and 
its image. The arguments of Section~6 shows that $\gamma_- = {1 \over (a - 1)h}$ can happen iff $a = 0$ and the maps are bipartite.

We now apply
the strategy of Section 3.3 in the non-generic critical case. The
trigonometric parametrization \vtrigpara\ reads here, for $a \neq 1$,
\eqn\trigpara{x(v) ={1\over (1-a^2)h}{(1-a)\cosh v+(1+a)\cosh v_\infty
\over \cosh v -\cosh v_\infty}\,,}
where
\eqn\coshinf{v_\infty = \arg\cosh
{(1-a)(1-(1+a)h\gamma_-)\over (1+a)(1+(1-a)h\gamma_-)}\ .}
The case $a=1$ is obtained by taking a suitable limit, which we will
discuss in the next section. Note that, since $\cosh
v_\infty$ is real, $v_\infty$ necessarily lies on the boundary of the domain
$[0,\infty[+{\rm i}[0,\pi]$.

Using this explicit parametrization, we may now determine the
resolvent exactly, by computing the coefficients in \omegagensol.
From now on, we will focus on the case where $g_k^{(0)}= g
\delta_{k,3}$, i.e. where faces not visited by the loops are
themselves triangles, weighted by $g$. We have in this case
$V'_0(x)=x-g x^2$ so that \condomega\ can be expanded into
\eqn\expanvinf{\eqalign{\omega(v)
 ={1\over (4-n^2)(1-a^2)^3}\Big\{&-6{A_3 {g\over h^3} \over (v-v_\infty)^4}
+2{A_2 {g\over h^3} +B_2 {1\over h^2} \over (v-v_\infty)^3}\cr 
& -{A_1 {g\over h^3}+B_1 {1\over h^2} 
\over (v-v_\infty)^2} +{C_0 \over (v-v_\infty)}+O(1)\Big\}\cr }}
with 
\eqn\ABC{\eqalign{
A_3 & =  {8 \coth^3 v_\infty \over 3}\cr
A_2 & = 8 \coth^2 v_\infty (a+\sinh^{-2} v_\infty)\cr
B_2 & = 4 \coth^2 v_\infty (1-a^2)\cr
A_1 & = {4 \coth v_\infty \over 3} \left( 6\coth^4 v_\infty +
      (6a-8) \coth^2 v_\infty + 3(1-a)^2 \right)\cr
B_1 & =  4 \coth v_\infty (1-a^2) \left(\coth^2 v_\infty - (1-a)\right) \cr
C_0 & = 2 (1-a^2)^3 (n-2)\ .\cr}}
Then, since the fundamental solution $\zeta$ in \zetab\ has a simple pole at $0$
with residue $1$, satisfying \expanvinf\ leads us immediately to the
desired expression for $\omega$:
\eqn\wsolbis{\eqalign{\omega(v)= &{1\over (4-n^2)(1-a^2)^3} \Big\{
A_3 {g \over h^3}  \left(\zeta'''(v-v_\infty) - \zeta'''(-v-v_\infty)\right)
\cr & \qquad +\left( A_2 {g\over h^3}+B_2 {1\over h^2}\right) 
 \left(\zeta''(v-v_\infty) - \zeta''(-v-v_\infty)\right) \cr & \qquad +
\left( A_1 {g\over h^3}+B_1 {1\over h^2}\right)
\left(\zeta'(v-v_\infty) - \zeta'(-v-v_\infty)\right) \cr & \qquad \qquad
\qquad \qquad +
C_0 \left(\zeta(v-v_\infty) - \zeta(-v-v_\infty)\right)\Big\}\ .\cr}}
In order for the solution \wsolbis\ to be consistent, we should ensure
that the spectral density $\rho$ vanishes at $\gamma_\pm$. 
As discussed in Section 3.3, demanding that $\rho(\gamma_-)=0$ amounts
to the condition
\eqn\condgammam{\omega({\rm i}\pi)=0}
where, from \wsolbis,
\eqn\condexpl{\eqalign{& \omega({\rm i}\pi)\propto
A_3 {g \over h^3}  
\left(
b \cosh (b v_\infty) \coth v_\infty \Big(b^2+6 \sinh^{-2}v_\infty\right)
\cr & \qquad \qquad \quad -
\big(2 \sinh^{-4}v_\infty+\left(3b^2+4\coth^2v_\infty\right)
\sinh^{-2}v_\infty+b^3\big)\sinh(b v_\infty)\Big)
\cr & +\left( A_2 {g\over h^3}+B_2 {1\over h^2}\right) 
 \left(
b\cosh(b v_\infty) \left(2 \sinh^{-2}\!v_\infty\!+\!b\right)\!-\!\coth v_\infty
\left(b^2\!+\!2 \sinh^{-2}\!v_\infty\right)\sinh(b v_\infty)\right) \cr & +
\left( A_1 {g\over h^3}+B_1 {1\over h^2}\right) \left( {1\over 2}
\sinh^{-2}v_\infty \left((b-2) \sinh(b v_\infty)- b \sinh
((b-2)v_\infty)\right) \right) \cr & +
C_0 
\sinh^{-1}v_\infty \sinh((1-b)v_\infty)\ .\cr}}
Similarly, demanding that $\rho(\gamma_+)=0$ amounts, as already discussed,
to the condition 
\eqn\condgammap{\omega(v)=o(e^{-v}) \quad {\rm for}\ v \to +\infty.}
By \zetasymp\ and \wsolbis, we have the asymptotic expansion
 \eqn\omegasymp{\omega(v) =\kappa(b) e^{-b v}+\kappa(2-b) e^{-(2-b)v}+
o(e^{-(2-b)v})}
with
\eqn\kappaval{\eqalign{\kappa(b)= &{1\over (4-n^2)(1-a^2)^3} \Big\{
A_3 {g \over h^3}(-2 b^3 \sinh (b v_\infty))
+\left( A_2 {g\over h^3}+B_2 {1\over h^2}\right) 
(2 b^2 \cosh (b v_\infty))
\cr & \qquad \qquad \qquad +
\left( A_1 {g\over h^3}+B_1 {1\over h^2}\right)
(-2 b \sinh (b v_\infty))
+C_0 
(2 \cosh (b v_\infty))
\Big\}\ .\cr}}
Then, the condition \condgammap\ reads nothing but 
\eqn\condgammapbis{\kappa(b)=0\ .}
Finally, the positivity of the $F_k$'s requires by \Fkdense\
the extra condition 
\eqn\condpos{\kappa(2-b)\geq 0.}

\subsec{Phase diagram}

\fig{Phase diagram of the $O(n)$ loop model with bending energy in
the $(g,h)$ plane. The various plots correspond to $b=0.3$ ($n=2 \cos \pi b$)
and, from top to bottom, $a=0.5$ (red), $1$ (green) and $2$ (blue). 
Plots for other values of $b$ and $a<a_c(n)$ would essentially have the same aspect.
The solid line corresponds to non-generic critical models and is {\it computed
exactly} upon solving eqs.\condgammam\ and \condgammapbis. In the absence of 
the third constraint  \condpos, this line would continue as indicated in dashed 
line (yellow). The condition \condpos\ imposes that the line terminates
at a point $(g^*,h^*)$. All points on the non-generic critical line
lie in the universality class of the dense $O(n)$ model except the endpoint 
$(g^*,h^*)$ which lies in the universality class of a dilute $O(n)$ model.
For the various models, a dot-dashed line was drawn (but not computed exactly 
here) to represent the line of generic critical points. This line connects 
the point $(g^*,h^*)$ to the point $({1\over 2\cdot 3^{3/4}},0)$ describing 
pure triangulations. The model is well-defined for values of $g$ and $h$ lying 
below the (generic or non-generic) critical lines. The non-generic critical line reduces
to a single dilute point with $g^*=0$ at $a=a_c(n)$, and vanishes for $a>a_c(n)$.}
{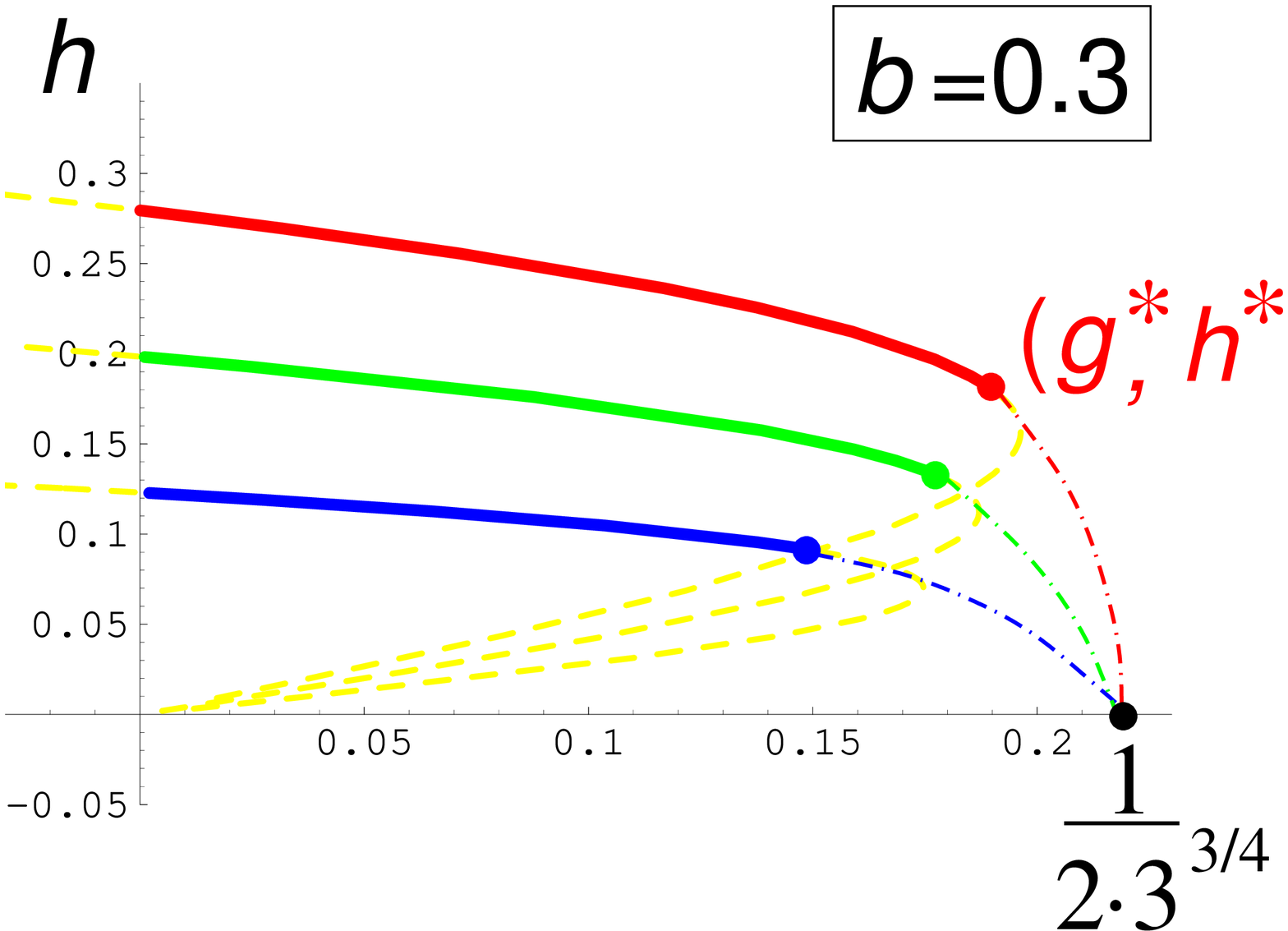}{11.cm}
\figlabel\phasediag

Fixing $a\geq 0$ and $b$ strictly between $0$ and $1/2$ (i.e. $0<n<2$), 
the consistency conditions \condgammam, \condgammapbis\ and \condpos\ determine
the values of $g$ and $h$ where the model is non-generic critical. 
Viewing $v_\infty$ as a parameter, \condgammam\ and \condgammapbis\
form a system of inhomogeneous linear equations for the variables 
$g/h^3$ and $1/h^2$ (or equivalently for the variables $g/h$ and $h^2$
after multiplying the equations by $h^2$). Solving this system, we obtain a 
family of non-generic critical points, forming a line in the $(g,h)$ plane
parametrized by $v_\infty$. Recall that $\cosh v_\infty$ is real, so
that $v_\infty$ varies on $[0,\infty[\cup ({\rm i}[0,\pi])\cup ([0,\infty[+
{\rm i}\pi)$. This range is further restricted by the requirement that 
$g\geq 0$ and $h\geq 0$, and by the condition \condpos. We do not give here 
the precise parametrization of the non-generic critical line as the formulas 
are quite cumbersome, but we instead present in Fig.~\phasediag\ the 
corresponding plots for $b=0.3$ and several values of $a$. Plots for other
values of $b$ and $a$ (not too large) have essentially the same aspect:
the line starts at 
$g=0$ and a finite value of $h$ and ends at a point $(g^*,h^*)$.
The value of $g^*$ decreases with increasing $a$ and reaches $0$ for a threshold
value $a=a_c(n)$. For $a>a_c(n)$, the non-generic critical line
disappears. The value of $a_c(n)$ decreases with increasing $n$
from $0$ to $2$, with the particular values
$a_c(0)=\infty$, $a_c(1)=4$ and $a_c(2)=2$.
The physical meaning of this phenomenon is under investigation.

For $a<a_c(n)$, all points on the non-generic critical line,
except the endpoint $(g^*,h^*)$, lie in the
universality class of the dense $O(n)$ model: in particular
$F_k$ has the asymptotic behaviour \Fkdense\ (with
$C'=\kappa(2-b)>0$). The endpoint $(g^*,h^*)$ lies instead in the
universality class of the dilute $O(n)$ model: $F_k$ has the
asymptotic behaviour \Fkdilute\ (with $C''=\kappa(2+b)<0$).
On the plots, we have also indicated the existence of a line
of generic critical points which links the point $(g^*,h^*)$ 
to the generic critical point of pure triangulations 
without loops, obtained for $h=0$ and $g=1/(2\cdot 3^{3/4})$.
An explicit parametric representation for this line can be also
obtained by the strategy of Section 2.5, but its expression is even
more cumbersome than that of the non-generic critical line (see for
instance [\xref\BBG, Section 6.4] for its computation in the $a=0$ case,
with different face weights of the form $g_k^{(0)}=g \delta_{k,4}$).

\fig{Plots of the spectral density of the $O(n)$ loop model with
bending energy at $b=0.3$ and for $a=0.5$, $1$ and $2$. For each case,
we plotted the spectral density at the non-generic critical dilute point
$(g^*,h^*)$ (solid line), and at some (somewhat arbitrary) non-generic 
critical dense point along the non-generic critical line 
(dashed line).}{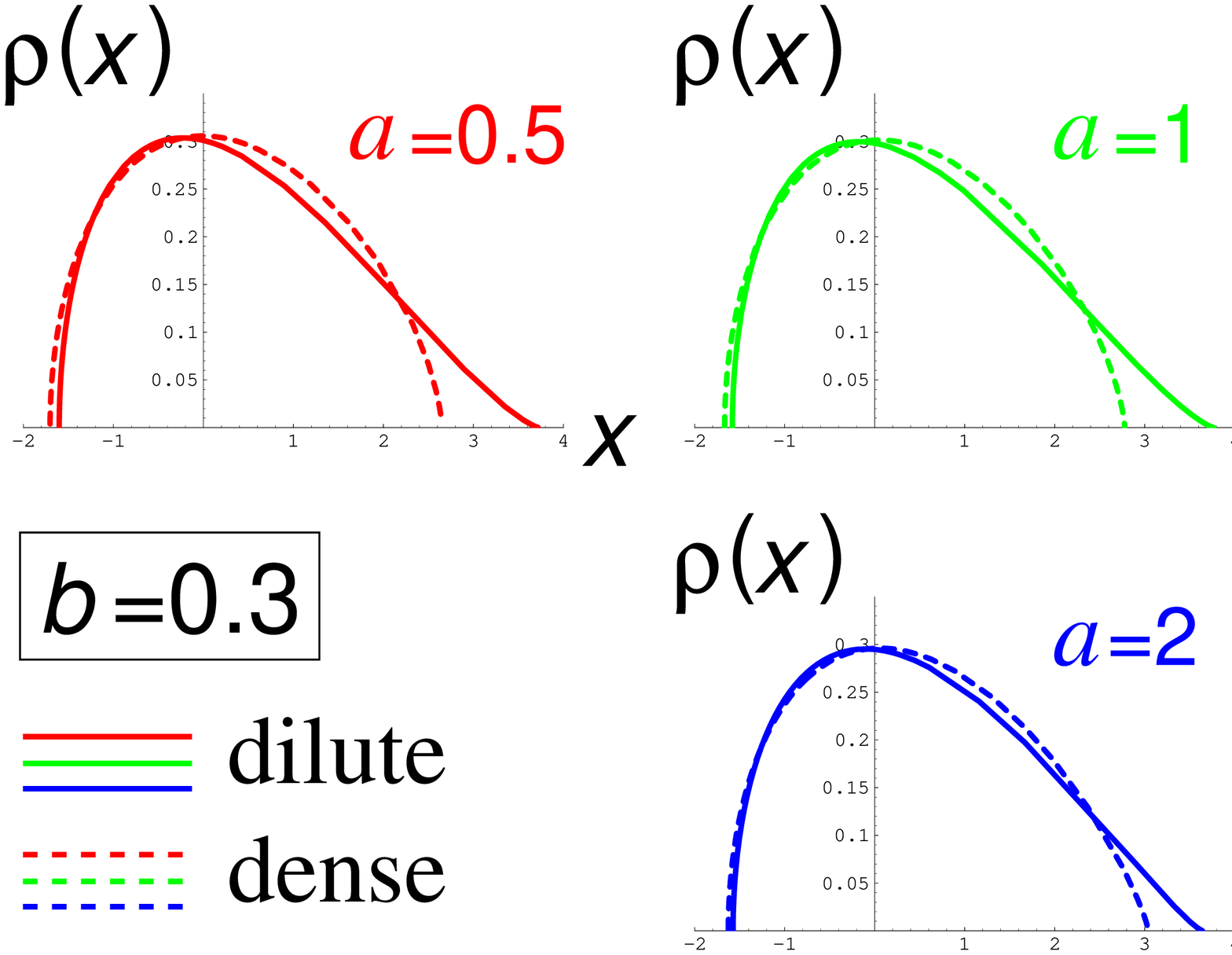}{12.cm}
\figlabel\spectral
For illustration, we have plotted in Fig.~\spectral\ the spectral density of eq.~\rhopara\ at 
the dilute point and at some (arbitrarily chosen) dense point on the
non-generic critical line for $b=0.3$ and for the three values of $a$ 
displayed in Fig.~\phasediag.

As already mentioned, the case $a=1$ is special. Still, the equation
of the non-generic critical line may in practice be obtained by 
simply taking the $a\to 1$ limit of the consistency conditions 
\condgammam, \condgammapbis\ and \condpos, with $v_\infty$ tending
simultaneously to ${\rm i} \pi/2$ as
\eqn\vinfaone{v_\infty={{\rm i}\pi\over 2}+{\rm i}(a-1)
\left({1\over 2} -h\,\gamma_-\right)+O\left((a-1)^2\right)}
so that Eq.~\coshinf\ remains satisfied. Solving the first two 
conditions in this limit leads 
to the following parametrization of the non-generic 
critical line
\eqn\ngclaone{\eqalign{{g\over h}& =
{2b\sqrt{2 + n}\,\rho - 2\sqrt{2 - n} \over 2b\sqrt{2 + n}\,\rho -\sqrt{2 - n}\big(1 + {1 - b^2 \over 2}\,\rho^2\big)}
 \cr h^2 & = 
{b\rho^2 \over 48\sqrt{4 - n^2}}\,{
b\sqrt{2 + n}\big(6 + (1 - b^2)\rho^2\big) 
-4(1 - b^2)\sqrt{2 - n} \rho\over 2b\sqrt{2 + n}\,\rho - \sqrt{2 - n}\big(1 + {1-b^2 \over 2}\rho^2\big)}\ ,\cr}}
where we use $\rho=1-2h\gamma_- = 1 - {\gamma_- \over \gamma_+} \geq 0$ as a parameter along the line. The corresponding plot is displayed in Fig.~\phasediag. From the third 
condition, we find that the line ends at the dilute point
\eqn\diluteaone{{g^*\over h^*} = 
1 + \sqrt{{2 - n \over 6 + n}},\qquad (h^*)^2= {(2 - b)b \over 12(1 - b)^2}\,{4 - \sqrt{(2 - n)(6 + n)} \over (2 + n)\sqrt{(2 -n)(6 + n)}}\ .}
These formulas are in agreement with the results of \Kost. In particular, 
we find the value $h= {1 \over 2\sqrt{2}\,\sqrt{2 + n}}$ for the non-generic critical 
point at $g=0$, with $\gamma_-= \sqrt{2}\sqrt{2 + n}\big(1 - {\sqrt{2 - n} \over b}\big)$ at this point.

For $n=1$ ($b=1/3$) and $a=1$, the model is equivalent to the Ising
model on the vertices of a random triangulation at zero magnetic
field, as seen by viewing the loops as domain walls for the Ising
spins. Besides the point $g^*={\sqrt{5} \over 2 \sqrt{2} \cdot
7^{3/4}}$, $h^* = {1 \over 12} \sqrt{{20 \over \sqrt{7}} - 5}$ which
corresponds to the transition point of the Ising model, another
interesting point on the non-generic critical line is found for
$g=h$. This point corresponds to having a vanishing Ising coupling in which
case the model reduces to Bernoulli-$1/2$ site percolation on random
triangulations: by \ngclaone\ we find $\rho=3/2$ and $g=h={1 \over 2
\sqrt{2} \cdot 3^{3/4}}$. Note that this value is equal to the
critical value of $g$ for pure triangulations divided by $\sqrt{2}$ as
it should, since the two choices of spin variables at each vertex are
equivalently counted by assigning a weight $\sqrt{2}$ per triangle
(there are essentially twice as many triangles as vertices in a planar
triangulation).

\subsec{The $n\to 0$ limit}

\fig{Exact phase diagram of the $O(n)$ loop model with bending energy in
the $(g,h)$ plane, in the limit $n\to 0$ and, from top to bottom, for
$a=0.5$ (red), $1$ (green) and $2$ (blue). This phase diagram is similar
to that of Fig.\phasediag, apart from the fact that the line of non-generic
critical points now has an infinite slope at its endpoint $(g^*, h^*)$.
We have $g^*={1\over 2\cdot 3^{3/4}}$ independently of $a$, so that the generic 
critical points now form a vertical segment (dashed line).}
{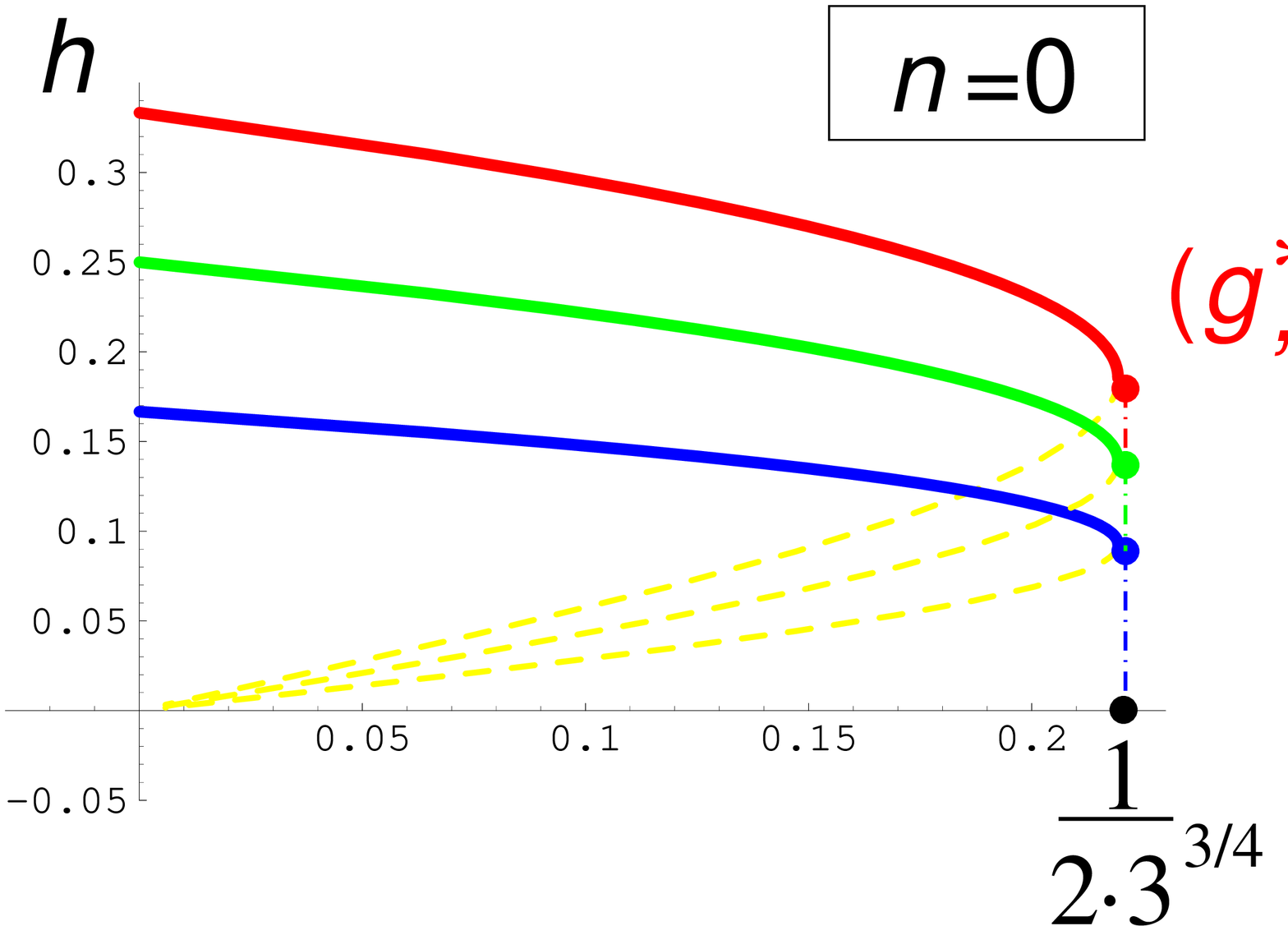}{11.cm}
\figlabel\phasediagzero

Fig.~\phasediagzero\ shows the phase diagram obtained by our calculation 
in the $n\to 0$ limit. In this limit, the equation for the non-generic 
critical line reads explicitly, in parametric form:
\eqn\nzero{\eqalign{{g\over h}& ={2(a^2-1)(c_{\infty} + 1)
(ac_\infty+ a-1) \over 2(a - 1)^2 + 4a(a - 1)c_{\infty} + (2a^2 + 1)c_{\infty}^2} \cr
h^2 & ={ -c_\infty^2 \over 4(a^2 - 1)^2(c_\infty+1)^2}\,{2(a - 1)^2 + 4a(a - 1)c_\infty + (2a^2 - 1)c_\infty^2 \over 2(a - 1)^2 + 4a(a - 1)c_\infty + (2a^2 + 1)c_\infty^2} .\cr}}
The parameter $c_{\infty} = \cosh v_{\infty}$ ranges a priori over the real line. By computing discriminants, we observe that the denominator in both fraction is always positive, while the numerator in the expression of $h^2$ has two reals roots at $c_{\infty} = -2{a - 1 \over a \pm 1/\sqrt{2}}$. As a consequence, one should reject the values of $c_\infty$ leading to negative values of ${g \over h}$ or $h^2$.

 The line ends at the dilute point
\eqn\nzerodilute{g^*={1\over 2\cdot 3^{3/4}}\ , \qquad
h^*= {1\over (1+a)} {1\over 3^{3/4}+3^{1/4}}}
and, since the value of $g^*$ matches that of the (generic) critical point of
pure triangulations, we expect that the generic critical line 
becomes the vertical segment $\{(g^*,h),0\leq h < h^*\}$.

Let us now explain how to recover this phase diagram without recourse to our
calculation by the following simple argument. The $n\to 0$ limit
corresponds to configurations of maps with a fixed finite number of loops.
The generating function for such configurations may be expressed 
in terms of the generating function for pure triangulations with a boundary,
gathered in the resolvent. For pure triangulations, 
this resolvent has a cut with endpoints $S\pm 2 \sqrt{R}$
where $R$ and $S$ satisfy the relations (see for instance \CENSUS)
\eqn\reltri{R=1+2 g R S, \qquad  S= g (2R+S^2).}

Now criticality may be achieved in two different ways. The loops may remain 
small, in which case a generic critical point is simply obtained whenever 
$dR/dg=\infty$, i.e. $g=g^*=1/(2 \cdot 3^{3/4})$ irrespectively of $h$ 
(small enough). We therefore deduce the existence of a vertical line of 
generic critical points in the $(g,h)$ plane at $g=g^*$.
On the other hand, non-generic criticality occurs when the loops become
large. Now a loop with contours of lengths $k$ and $k'$ yields a 
contribution of order $F_k^{{\rm pure\ tri}}F_{k'}^{{\rm pure\ tri}}
A_{k,k'}$ to the partition function. In particular, loops of length
$K$ (i.e. with $k+k'=K$) contribute at large $K$ as 
\eqn\contestim{(S+2\sqrt{R})^K \sum_{k+k'=K} A_{k,k'}\simeq
(S+2\sqrt{R})^K\left((a+1)h)\right)^K}
(the reader may easily derive the identity $\sum_{k+k'=K} A_{k,k'}
=(x_+^{-K}+x_-^{-K})/2$ where $x_\pm =1/((a\pm 1)h)$ are the fixed points
of the involution). The condition for the occurrence of large loops 
reproduces precisely the non-generic critical condition \coales, namely
\eqn\coalestri{\gamma_{+}={1\over (a+1)h}=S+2\sqrt{R}\ .}
Using \reltri, this leads immediately to the following parametric form of the 
non-generic critical line:
\eqn\paramsigma{\eqalign{{g\over h}& = (1+a) 
\left(\sqrt{2 \sigma(1-\sigma)} +\sigma\right)\cr 
g^2 & = {\sigma(1-\sigma)(1-2\sigma) \over 2}\cr}}
where we have set $\sigma= g S$ (this parametrization has been 
introduced in \CENSUS). The value of the parameter $\sigma$ 
varies along the line between $0$ and $\sigma^*=(3-\sqrt{3})/6$, 
corresponding to the dilute point \nzerodilute, which is also the endpoint of 
the vertical line of generic critical points.
Now it is a straightforward exercise to check that the parametrizations
\nzero\ and \paramsigma\ are indeed fully equivalent upon relating 
the varying parameters $\sigma$ and $c_\infty$ through 
\eqn\valsigma{\sigma= {2(ac_\infty+a-1)^2\over 2(a - 1)^2 + 4a(a - 1)c_{\infty} + (2a^2 + 1)c_{\infty}^2}\ .}

\newsec{The twisting loop model on quartic maps}

The analysis of Section 4 can be applied to obtain the non-generic critical line
of the twisting loop model defined in Section 2. To be explicit, 
we decide to set $g^{(0)}_k=g \delta_{k,4}$, i.e. the faces not visited
by the loops are themselves squares, weighted by $g$. Eq.~\cuttwist\
reduces in this case to 
\eqn\cuttwistquad{\tilde{W}(X+{\rm i}0)+\tilde{W}(X-{\rm i}0)= 
1 -g X - 2 n\, \tilde{W}(h_2^{-1} - X)\ ,}
of which we look for a solution $\tilde{W}(X)$ having a cut 
along the real segment $[0,\Gamma]$ for some positive $\Gamma$
(recall that the resolvent $W(x)$ is recovered via 
$W(x)=x\, \tilde{W}(x^2)$ and has a cut along the real segment 
$[\gamma_-,\gamma_+]$, with $\gamma_\pm=\pm\sqrt{\Gamma}$).
Equation \cuttwistquad\ is similar to (yet even simpler than) \funceq\
and we may solve it by the strategy of Section 3.3. It now involves
the particular involution $X \mapsto s(X)=h_2^{-1}-X$, here $n$ is
replaced by $2n$ and the polynomial $V'_0$ reads $\tilde{V}_0'(X)=1-g X$.
As before, we write
$\tilde{W}(X)=\tilde{W}_{{\rm part}}(X)+\tilde{W}_{{\rm hom}}(X)$,
with the particular solution (with no cut)
\eqn\ciswpart{\tilde{W}_{{\rm part}}(X)=
{(1-g X)-n\left(1-g(h_2^{-1}-X)\right)\over 2(1-n^2)}}
and where $\tilde{W}_{{\rm hom}}(X)$ is now a solution of the homogeneous 
equation
\eqn\ciswhomeq{\tilde{W}_{{\rm hom}}(X+{\rm i}0)+\tilde{W}_{{\rm hom}}
(X-{\rm i}0)+2n\, \tilde{W}_{{\rm hom}}(h_2^{-1}-X)= 0\ .}
Again, among all solutions of this equation, $\tilde{W}_{{\rm hom}}(X)$ is 
determined by requiring that $\tilde{W}(X)$ has no poles at finite 
values of $X$ and that $\tilde{W}(X) \sim 1/X$ for $X \to \infty$. 
We are here in a situation where $s(\infty)=\infty$ so that, as
for loops on triangulations with $a=1$, we have a single condition 
to satisfy, namely
\eqn\ciscond{\tilde{W}_{{\rm hom}}(X)=-{(1-g X)-n (1-g(h_2^{-1}-X))\over 
2(1-n^2)} +{1\over X} +O\left({1\over X^2}\right)\ .}

The general solution of \ciswhomeq\ is still given by elliptic
functions. Note however that in the elliptic integral \vint, $\gamma_+$
and $\gamma_-$ should be respectively replaced by $\Gamma$ and $0$. We
again concentrate on the non-generic critical case which occurs
whenever the endpoint $\Gamma$ of the cut is a fixed point of the
involution $s(X)$, namely $\Gamma=1/(2 h_2)$. The parametrization
\vtrigpara\ is here replaced by
\eqn\cistrigparam{X(v)={1\over 2h_2}\left(1+{1\over\cosh v}\right).}
Defining
\eqn\cisomega{\tilde{\omega}(v)=X'(v)\tilde{W}_{{\rm hom}}(X(v))}
as in \whompara, we find that $\tilde{\omega}(v)$ is an odd
meromorphic function satisfying the functional equation
\eqn\cisomegaeq{\tilde{\omega}(v+{\rm i}\pi)+\tilde{\omega}(v-{\rm i}\pi)-
2n\, \tilde{\omega}(v) = 0\ ,}
while Eq.~\ciscond\ translates into
\eqn\ciscondbis{\tilde{\omega}(v) =
{g\over 8 h_2^2 (1-n)}{1\over (v-{{\rm i} \pi\over 2})^3}
+ {{\rm i} (g-2h_2)\over 8 h_2^2 (1+n)}{1\over (v-{{\rm i} \pi\over 2})^2}
-{1\over v-{{\rm i} \pi\over 2}}+O(1)\ .}
Again, $\tilde{\omega}(v)$ may be expressed in terms of the
fundamental solution $\zeta$ of Eq.~\zetab, but now with
$b$ given by 
\eqn\newnbrel{\pi b = \arccos n}
which lies in the range $0<b<1/2$ for $0<n<1$. 
We get explicitly
\eqn\ciswsol{\eqalign{\tilde{\omega}(v)& =
{g\over 16 h_2^2 (1-n^2)}\left(\zeta''\left(v-{{\rm i}\pi\over 2}\right)
- \zeta''\left(-v-{{\rm i}\pi\over 2}\right)\right)\cr &  \qquad -
{{\rm i} (g-2h_2)\over 8 h_2^2 (1-n^2)}
\left(\zeta'\left(v-{{\rm i} \pi\over 2}\right)
-\zeta'\left(-v-{{\rm i}\pi\over 2}\right)\right) \cr & 
\quad \qquad - {1\over 1+n}
\left(\zeta\left(v-{{\rm i}\pi\over 2}\right) -
\zeta\left(-v-{{\rm i}\pi\over 2}\right)\right)\ .\cr}}
Since $W(x)=x\, \tilde{W}(x^2)$, the spectral density is recovered
via 
\eqn\cisrho{\rho(x)={|x|\over 2{\rm i}\pi}
\left(\tilde{W}_{{\rm hom}}(x^2 + {\rm i}0)-\tilde{W}_{{\rm hom}}(x^2-{\rm i} 0)
\right)}
so that
\eqn\cisrhoparam{\rho\left(\pm \sqrt{X(v+{\rm i}\pi)}\right)=
{\sqrt{X(v+{\rm i}\pi)}\over X'(v+{\rm i}\pi)}\cdot
{\tilde{\omega}(v+{\rm i}\pi)-\tilde{\omega}(v-{\rm i}\pi)\over 
2 {\rm i}\pi}}
for real positive $v$. Contrary to the previous case, we need not have
$\tilde{\omega}({\rm i}\pi)=0$, since $\rho(0)$ does not
vanish. But we must impose again the condition $\tilde{\omega}(v)=o(e^{-v})$ 
for $v\to \infty$ so that $\rho(\gamma_{\pm})=0$. 
By \zetasymp\ and \ciswsol, we have the expansion
\eqn\cisomegasymp{\tilde{\omega}(v)=\tilde{\kappa}(b)e^{-b v}+
\tilde{\kappa}(2-b) e^{-(2-b) v}+o(e^{-(2-b) v})}
with
\eqn\tildekappa{\tilde{\kappa}(b)= {(g\, b^2+16\, h_2^2\, (n-1))\,
\cos{\pi b\over 2} -2\,b\,(g-2\,h_2)\,\sin{\pi b \over 2}
\over 8\, h_2^2\,(1-n^2)}\ .}
The condition $\tilde{\kappa}(b)=0$ yields
\eqn\cisparabola{g={4 b\,h_2-16\sqrt{1 - n^2}\,h_2^2 \over 
b\left(2-b\sqrt{{1 + n \over 1 - n}}\right)}\ .}
Finally, as before, the positivity of $F_k$ requires
that $\tilde{\kappa}(2-b) \geq 0$, which yields the condition 
\eqn\cisparaposcond{g\leq {4\,(2-b)\,h_2+16\sqrt{1 - n^2}\,h_2^2\over 
(2-b)\left(2+(2-b)\sqrt{{1 + n \over 1 - n}}\right)}\ .}
The non-generic critical line therefore ends at the dilute point
\eqn\cisdilute{\eqalign{g^*& = 
{b (2-b)\over 4(\sqrt{1-n}+ (1 - b)\sqrt{1 + n})^2}\ ,
\cr h_2^* & ={b(2-b)\over 8 (1- n +(1-b)\sqrt{1 - n^2})}
\ .\cr}}
\fig{The phase diagram of the twisting loop model in the variables
$(g,h_2)$ for $b=0.3$ ($n=\cos\pi b$). The line of (dense) non-generic critical points 
(solid red line) is the portion of the parabola \cisparabola\ (dashed yellow)
delimited by the conditions $g\geq 0$ and \cisparaposcond. The
line ends at the dilute point $(g^*,h_2^*)$ of eq.~\cisdilute. We indicated
in dot-dashed line the existence of a line of generic critical points
linking the point $(g^*,h_2^*)$ to the point $(1/12,0)$ describing
pure quadrangulations.}{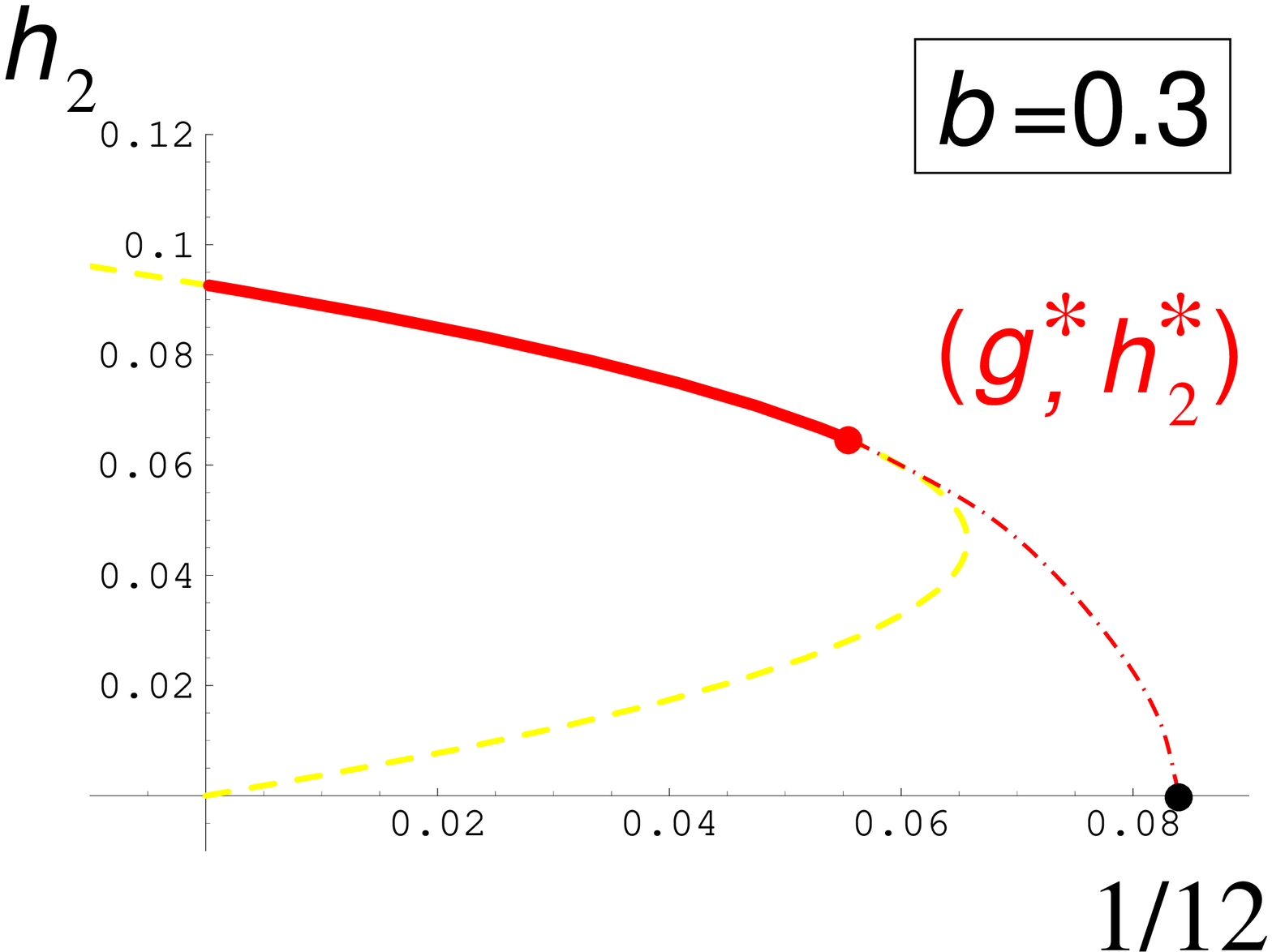}{11.cm}
\figlabel\phasediagcis
\fig{Plots of the spectral density of the twisting loop model at $b=0.3$.
We plotted the spectral density at the non-generic critical dilute point
$(g^*,h_2^*)$ of eq.~\cisdilute\ (solid line), and at some 
(arbitrary) non-generic critical dense point along the non-generic critical 
line (dashed line).}{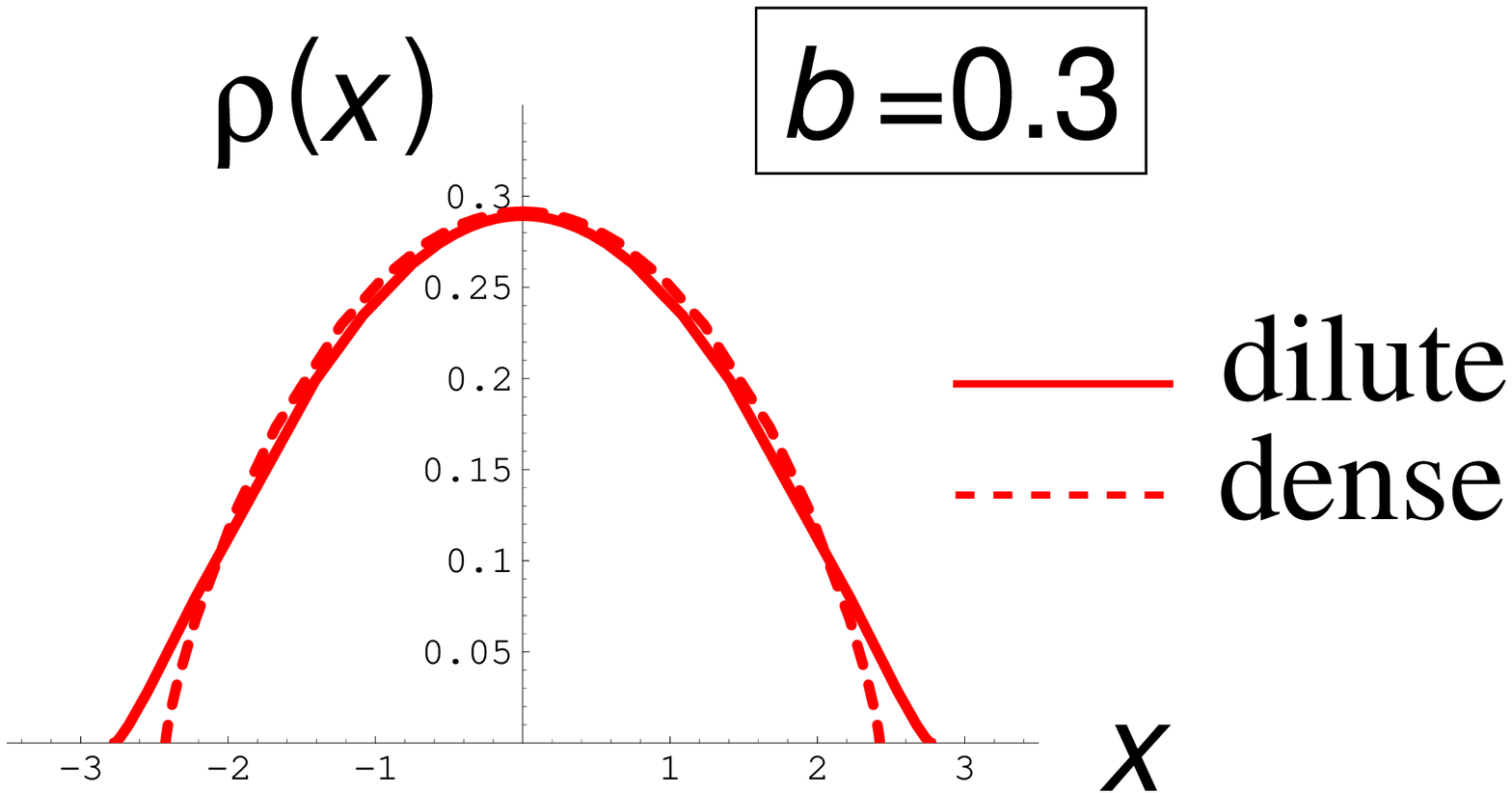}{10.cm}
\figlabel\spectralcis
The phase diagram of the twisting loop model for $b=0.3$ 
is displayed in Fig.~\phasediagcis.
A line of generic critical points links the dilute point $(g^*,h_2^*)$ 
to the critical point of pure quadrangulations without loops at $g=1/12$, 
$h_2=0$. For illustration, we have also plotted in Fig.~\spectralcis\ 
the spectral density at the dilute point and at some (arbitrarily chosen) 
dense point on the non-generic critical line.
\fig{Phase diagram of the twisting loop model at $n=0$.
The line of non-generic critical points now has an infinite slope at its 
endpoint $(g^*, h_2^*)$. We have $g^*={1/12}$ so that 
the generic critical points now form a vertical segment (dashed line).}{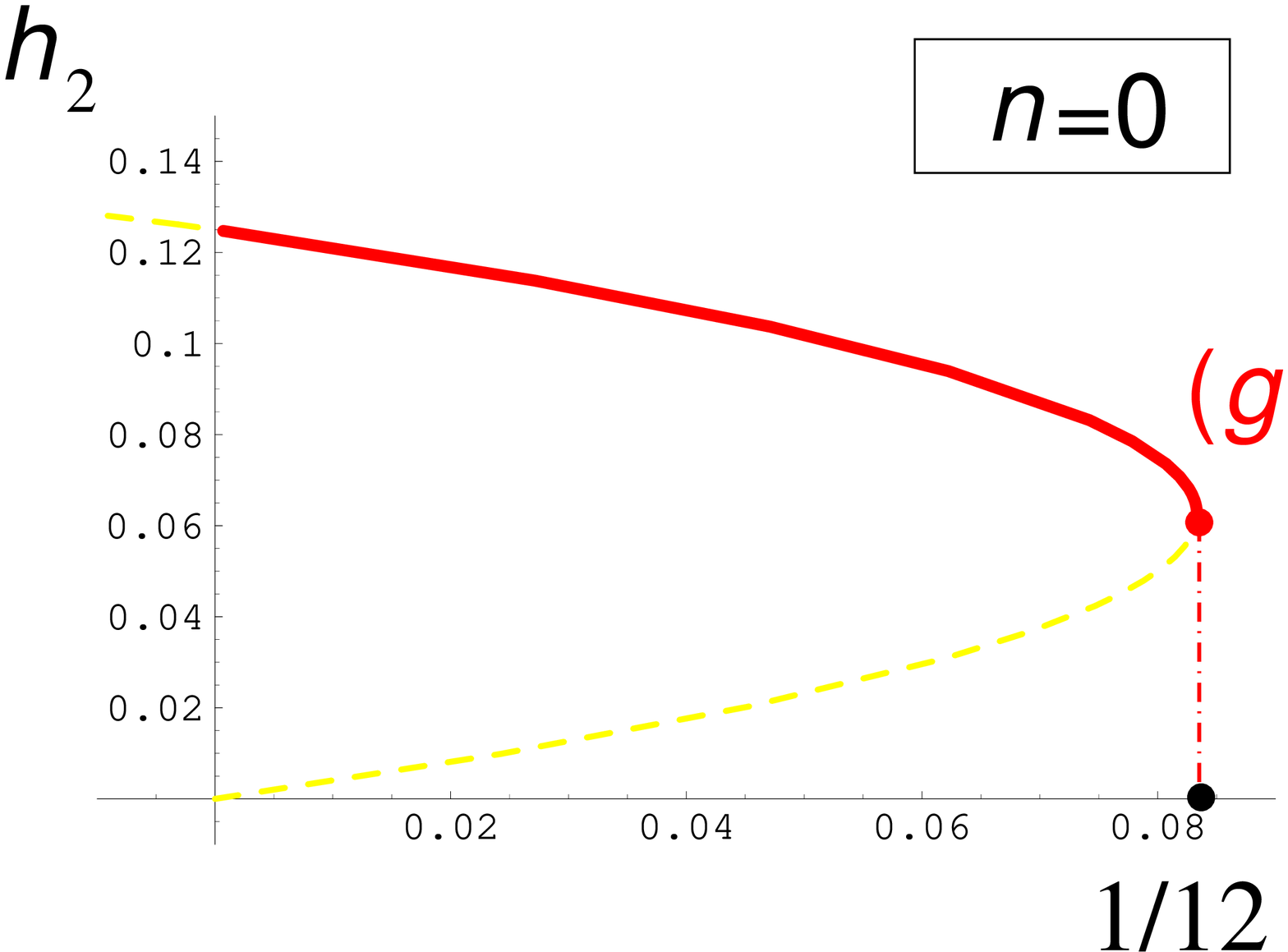}{11.cm}
\figlabel\phasediagciszero
When $n\to 0$, the dilute point \cisdilute\ has coordinates $g^*=1/12$ 
and $h_2^*=1/16$, with the same value of $g$ as for critical pure 
quadrangulations, so that the generic critical line becomes a 
vertical segment $\{(1/12,h),0<h<1/16\}$ (see Fig.~\phasediagciszero). 
As for the equation for the 
the non-generic critical line, it may be recovered by simply writing that
$\Gamma=(\gamma_+)^2$ is the fixed point $1/(2h_2)$ of the involution, with, 
for $n\to 0$, a value of $\gamma_+$ equal to that of pure quadrangulations. 
More precisely, we have as in Section 3.5 $\gamma_+=S+2\sqrt{R}$, 
but now with $S=0$ and $R$ solution of (see for instance \CENSUS)
\eqn\eqforR{R=1+3\, g\, R^2\ .}
Writing $(2\sqrt{R})^2=1/(2 h_2)$ leads to 
\eqn\ciseqzero{g={8\over 3} (h_2- 8 h_2^2)}
which matches \cisparabola\ specialized to $n = 0$, i.e. $b=1/2$.

\newsec{Analytic properties of the resolvent for maps with controlled face degrees}
In this Section, we review a number of analytic properties of the resolvent 
for maps with controlled face degrees, as we have used them in Section 2 in
the framework of our loop models. 
\subsec{The one-cut lemma for maps with controlled face degrees}

We say that a sequence of non-negative numbers $(g_k)_{k \geq 1}$ is {\it admissible} 
when the generating function of pointed rooted planar maps with weight $g_k$ per face
of degree $k$ is finite. In particular, a model with bounded face degrees corresponds
to having all but a finite number of $g_k$'s non-zero, and such a sequence is admissible 
when these weights are small enough. 

We denote as before by ${\cal F}_p(g_1,g_2,\ldots)$ the multivariate generating function for 
maps with a boundary of length $p$ (with the convention ${\cal F}_0=1$),
which is finite whenever the sequence $(g_k)_{k\geq 1}$ is admissible, and we 
introduce the resolvent
\eqn\defreso{{\cal W}(x) = \sum_{p \geq 0} {{\cal F}_p(g_1,g_2,\ldots) \over x^{p + 1}}}
(the term resolvent, and the convention of taking ${\cal W}(x)$ as a series
in $1/x$ are borrowed from random matrix theory). 
The following property is essential to 
the determination of ${\cal W}(x)$ in specific models:
\medskip
\noindent $\bullet$ {\bf The one-cut lemma:} 
When $(g_k)_{k\geq 1}$ is an admissible sequence, there exists a real segment $I = [\gamma_-,\gamma_+]$ 
such that $x \mapsto {\cal W}(x)$ defines an analytic function on ${\bf C}\setminus I$, 
which has a finite discontinuity on $I$. Moreover, we have $|\gamma_-| \leq \gamma_+$ with equality iff 
$g_{l} = 0$ for all odd $l$. Lastly, the spectral density $\rho$, defined for $x \in I$ by
\eqn\rhodef{\rho(x) = {{\cal W}(x - {\rm i}0) - {\cal W}(x + {\rm i}0) \over 2{\rm i}\pi}\ ,}
vanishes at the endpoints of $I$, and is positive in the interior of $I$.

\medskip  This lemma essentially follows from the discussion in \BMJ\
for the case of a model with bounded face degrees, and we shall extend it to the case of
unbounded degrees below. Before, let us review the possible singular behaviours
of ${\cal W}$ at the dominant singularity $\gamma_+$
($\gamma_-$ being subdominant except in the bipartite case where ${\cal W}$ is odd, hence
has the same behaviour at both singularities).
For a generic admissible weight sequence, ${\cal W}$ has a
square-root singularity which, by transfer, corresponds to having ${\cal F}_p \sim C\, \gamma_+^p/p^{3/2}$ for $p \to \infty$. We say that the
 model is {\it non-critical} in this case. Upon increasing the weights up to a point
 where the sequence is on the verge of becoming non-admissible, we may reach
 a {\it critical} point where ${\cal W}$ has a higher-order singularity. Generically 
 (for instance, in any model with bounded face degrees), this singularity is of
 order $(x-\gamma_+)^{3/2}$, in which case ${\cal F}_p \sim C\,
 \gamma_+^p/p^{5/2}$ and we say, following the terminology of \BBG,
 that the model is {\it generic critical}. However, by suitably fine-tuning
 all the weights, we may obtain instead a {\it non-generic critical} model where
 the singularity is of intermediate order, for instance 
 $(x-\gamma_+)^{\beta}$ with $1/2<\beta<3/2$, corresponding to ${\cal F}_p
\sim C\,\gamma_+^p/p^{\beta+1}$. These cases encompass all the possible
singular behaviours for a non-negative admissible weight sequence.
 
Our derivation of the one-cut lemma relies on a description of the
discontinuity set of ${\cal W}$.
We use the relations established in \HANKEL\ thanks to the bijection between maps and mobiles, and the 
relation of the latter to Motzkin paths. We need some notations: first, let ${\cal F}_p[u]$ 
(resp. ${\cal F}_p^{\bullet}[u]$) be the generating function of rooted (resp. pointed rooted) maps with 
root degree $p$ where, on top of the face weights, each vertex of the map is counted with a weight $u$. 
Since the two definitions differ only by the marking of a vertex, we have
\eqn\mark{{\cal F}_p^{\bullet}[u] = u{d{\cal F}_p[u] \over du}\ .}
In other words
\eqn\FkFku{{\cal F}_p(g_1,g_2,\ldots) = {\cal F}_p[u = 1] = \int_0^1 {du \over u}\,{\cal F}_p^{\bullet}[u]\ .}
Then, let $P_p(R,S)$ be the generating function of lattice paths in ${\bf Z}^2$ from $(0,0)$ to $(p,0)$, 
where the set of allowed steps consists of:
\item{-} level-steps $(i,j) \rightarrow (i + 1,j)$ counted with weight $S$;
\item{-} up-steps $(i,j) \rightarrow (i + 1,j + 1)$ counted with weight $\sqrt{R}$;
\item{-} down-steps $(i,j) \rightarrow (i + 1,j - 1)$ also counted with weight $\sqrt{R}$.

\noindent An easy counting argument shows that  
\eqn\PkRS{P_p(R,S) = \sum_{i = 0}^{\lfloor p/2 \rfloor} {p! \over (i!)^2(p - 2i)!}\,R^i S^{p - 2i}\ .}
We may pack those numbers in the generating function 
\eqn\PxiRS{\widehat{P}(x,R,S) = 
\sum_{p \geq 0} {P_p(R,S) \over x^{p + 1}}=
{1 \over\sqrt{x^2 - 2S\, x + S^2 - 4R}}\ .}
We observe that there exists a segment $\widetilde{I}(R,S) = [S - 2\sqrt{R},S + 2\sqrt{R}]$ such 
that $x \mapsto \widehat{P}(x,R,S)$ is analytic in ${\bf C}\setminus 
\widetilde{I}(R,S)$ and has a discontinuity on $\widetilde{I}(R,S)$.

The relation established in \HANKEL\ is that ${\cal F}_p^{\bullet}[u] = u\,P_p(R[u],S[u])$, 
where $R[u]$ and $S[u]$ are the unique power series in $u$ and the $g_k$'s determined by the system
\eqn\relRS{\eqalign{S[u] & = \sum_{k \geq 1} g_k\,P_{k - 1}(R[u],S[u]) \cr  
R[u] & = u - {S^2[u] \over 2} + {1 \over 2}\sum_{k \geq 1} g_k\,P_k(R[u],S[u])\ .}}
Actually, $R[u]$ and $S[u]$ have themselves combinatorial interpretations in terms
of maps, and $2R[u]+S^2[u]$ is the generating function for pointed rooted maps.
In particular, whenever the sequence $(g_k)_{k\geq 1}$ is admissible,
$R[u]$ and $S[u]$ are continuous increasing functions of $u\in [0,1]$,
satisfying $R[0]=S[0]=0$, and having a positive derivative for $u\in [0,1[$
(their derivative at $u=1$ are possibly infinite, which corresponds to a critical case).
From \defreso\ and \FkFku, the resolvent can be written as
\eqn\eqnsprep{{\cal W}(x) = \int_0^1 {du \over \sqrt{x^2 - 2S[u]\, x + S^2[u] - 4R[u]}}\ .}
We deduce that ${\cal W}(x)$ is an analytic function outside the discontinuity set
\eqn\Disco{I = \bigcup_{u \in [0,1]} \big[S[u] - 2\sqrt{R[u]},S[u] + 2\sqrt{R[u]}\big].}
It is clear that $u\mapsto S[u] + 2\sqrt{R[u]}$ is a strictly increasing function of $u$
whose derivative does not vanish. Although it is not obvious, we justify in Section 6.3 that
$u \mapsto S[u] - 2\sqrt{R[u]}$ must be strictly decreasing with a non-vanishing derivative. Hence
\eqn\Ide{I = [\gamma_-,\gamma_+]}
where
\eqn\edgecut{\gamma_- = S[1] - 2\sqrt{R[1]},\qquad \gamma_+ = S[1] + 2\sqrt{R[1]}}
In particular, we have $|\gamma_-| \leq \gamma_+$, with equality if and only if $S[1]=0$. 
From \relRS, this is equivalent to having
$g_{l} = 0$ for all odd integers $l$, which means that only faces of even degree are allowed. By \eqnsprep\ and by monotonicity of $u\mapsto S[u] \pm 2\sqrt{R[u]}$, the spectral density reads, for $x \in ]\gamma_-,\gamma_+[$,
\eqn\foi{\rho(x) = {1 \over \pi}\int_{u(x)}^1 {du \over \sqrt{4R[u]-S^2[u]+2S[u]\, x-x^2}}}
where $u(x)$ is obtained via $S[u(x)] + 2\sqrt{R[u(x)]} = x$ for $x
\geq 0$ and via $S[u(x)] - 2\sqrt{R[u(x)]} = x$ for $x < 0$. Clearly,
$\rho$ is positive on $]\gamma_-,\gamma_+[$ and the non-vanishing of
$(d/du)(S[u]\pm 2\sqrt{R[u]})$ ensures that $\rho$ is continuous and
vanishes as $x \to \gamma_\pm$, as announced. In addition,
when $(d/du)(S[u] + 2\sqrt{R[u]})$ is finite at $u=1$, we may see from \eqnsprep\ that ${\cal W}$
develops a square-root singularity at $\gamma_+$, i.e. the model is non-critical.
When this derivative is infinite, the model is critical: in particular, generic criticality
corresponds to $S[u] + 2\sqrt{R[u]}$ having itself a square-root singularity at $u=1$.

\subsec{Functional relation for maps with controlled face degrees}

We now assume that $(g_k)_{k\geq 1}$ is an admissible sequence, so that the 
one-cut lemma applies. The discontinuity of ${\cal W}$ on its cut is related 
to the spectral density $\rho$ by \rhodef\ and conversely, ${\cal W}$ is the 
Stieltjes transform of $\rho$, namely
\eqn\rhotoW{{\cal W}(x) = \int_{\gamma_-}^{\gamma_+} {\rho(y) dy \over x - y},
\qquad {\cal F}_k = \int_{\gamma_-}^{\gamma_+} y^k \rho(y) dy.}
We claim that the resolvent ${\cal W}(\xi)$ is characterized by 
${\cal W}(x) \sim 1/x$ and by
\eqn\cutone{\forall x \in ]\gamma_-,\gamma_+[\qquad {\cal W}(x+{\rm i}0) 
+ {\cal W}(x-{\rm i}0) = V'(x)\ ,}
where $V$ is the so-called ``potential'' defined via
\eqn\defV{V'(x) = x - \sum_{k \geq 1} g_kx^{k - 1}\ .}
In terms of the spectral density, this characterization is equivalent to 
$\int_I \rho(x)dx = 1$ and
\eqn\pvp{\forall x \in ]\gamma_-,\gamma_+[\qquad {\rm p.v.} 
\int {\rho(y)dy \over x - y} = {V'(x) \over 2}\ .}
In this paragraph, we shall justify this functional relation for 
general admissible sequences.

Tutte's recursive decomposition \TUTEQ\ of rooted planar maps allows to 
write, for any sequence of face weights,
\eqn\TutteA{{\cal F}_m = \sum_{k = 0}^{m - 2} {\cal F}_k{\cal F}_{m - 2 - k} 
+ \sum_{k \geq 1} g_k\,{\cal F}_{m + k - 2}\ .}
If we want to write this relation in terms of the resolvent ${\cal W}(x)$, we 
first need to know the location of the singularities of $V'(x)$ which, at first, 
only makes sense as a power series in $x$. From \relRS\ and the fact that 
$S[1]$ and $R[1]$ are finite for admissible sequences, we can deduce a lower bound on the 
radius of convergence of $V'(x)$. We use the asymptotic behaviour of $P_k(R[1],S[1])$ 
when $k \rightarrow \infty$, which can be extracted from \PxiRS\ by transfer 
theorems. Since $|\gamma_-| \leq \gamma_+$, $\gamma_+$ is always the dominant 
singularity of $\widehat{P}(x,R[1],S[1])$ and $P_k(R[1],S[1])=O(k^{-1/2}\gamma_+^{-k})$
(when $S[1]=0$, $\gamma_{-}=-\gamma_{+}$ and $P_k$ vanishes for odd $k$ but
the same estimate holds for even $k$). Accordingly, the radius of convergence of $V'$ is at 
least $\gamma_+$. 
In particular, $x \mapsto V'(x)$ defines a real-analytic function at 
least on the segment $]\gamma_-,\gamma_+[$. So, we may represent, for any 
integer $m \geq 1$
\eqn\ReP{\eqalign{-\delta_{k,2}{\cal F}_{m} + g_k\,{\cal F}_{m + k -2} & = 
\mathop{{\rm Res}}_{\xi \rightarrow \infty} \xi^{m - 1}\,(\delta_{k,2}-g_k)
\xi^{k - 1}\,{{\cal F}_{m + k - 2} \over \xi^{(m + k - 2) + 1}} \cr & =
\mathop{{\rm Res}}_{\xi \rightarrow \infty} \xi^{m - 1}(\delta_{k,2}-g_k)\xi^{k - 1}
\,{\cal W}(\xi) \cr & ={1 \over 2{\rm i}\pi}\oint_{{\cal C}} 
d\xi\,\xi^{m - 1}(\delta_{k,2}-g_k)\xi^{k - 1}\,{\cal W}(\xi)}}
where the contour ${\cal C}$ surrounds the cut $[\gamma_-,\gamma_+]$ in 
positive orientation, and is included in the domain of analyticity of $V'$. 
To be precise, if $V'$ happens to have a singularity at $\xi = \gamma_+$, 
we choose a contour sticking to this point when going around the cut. 
The same caution might be necessary at $\xi = \gamma_- = -\gamma_+$ in the 
case of maps with even face degrees. This allows to do the summation over $k$,
namely
\eqn\RePB{\sum_{k \geq 1}\eqalign{-\delta_{k,2}{\cal F}_m + 
g_k\,{\cal F}_{m + k -2} = - {1 \over 2{\rm i}\pi}\oint_{{\cal C}} 
d\xi\,\xi^{m - 1}\,V'(\xi){\cal W}(\xi)}\ .}
Now, let us take $x \in {\bf C}$ such that $|x| > \gamma_+$, and sum over 
non-negative integers $m$ the equations \TutteA\ with a weight $x^{-m}$. 
All the series involved are convergent and we obtain
\eqn\Quadrel{{\cal W}^2(x) - {1 \over 2{\rm i}\pi} \oint_{\cal C} 
{d\xi \over x - \xi}\,V'(\xi)\,{\cal W}(\xi) = 0\ .}
This is an equality between functions of $x$ which are analytic in the 
unbounded connected component of ${\bf C}\setminus{\cal C}$, so it must 
be valid for any $x$ in this domain. 
In particular, we can squeeze ${\cal C}$ to stick to the cut $I$, and compute 
the discontinuity of the left hand side of \Quadrel\ at a point $x$ interior 
to $I$, which should equal zero. We find the desired result
\eqn\cutoneB{\forall x \in ]\gamma_-,\gamma_+[\qquad {\cal W}(x + {\rm i}0) + 
{\cal W}(x - {\rm i}0) - V'(x) = 0\ .}
When $|\gamma_-| < |\gamma_+|$, this relation also holds at $x = \gamma_-$.

\subsec{Edge behaviour of the spectral density}

We now justify that $u \mapsto S[u] - 2\sqrt{R[u]}$ is a decreasing function of $u$ with a non-vanishing derivative. We may eliminate the case $S[u]=0$ (i.e. we assume that at least one 
$g_l$ for odd $l$ is non-zero) as the property is obvious in this case. 
Our proof, as presented here, relies on real analysis rather than 
combinatorics. We introduce the shortcut notations $T[u] = \sqrt{R[u]}$,
which is a strictly increasing function of $u$, and $U(\xi) = \sum_{k \geq 1} g_k\,\xi^{k - 1}$. By the same trick that we used in Section 6.2, the relations \relRS\ can be rewritten
\eqn\relRSV{\eqalign{S[u] & = {1 \over 2{\rm i}\pi}\oint_{{\cal C}} 
{d\xi\,U(\xi) \over \sqrt{\xi^2 - 2S[u]\xi + S^2[u] - 4T^2[u]}}\ , \cr
2T^2[u] + S^2[u] & = 2u + {1 \over 2{\rm i}\pi}\oint_{{\cal C}} 
{d\xi\,\xi\,U(\xi) \over \sqrt{\xi^2 - 2S[u]\xi + S^2[u] - 4T^2[u]}}\ .}}
These relations are also well-known, and we have justified them 
for arbitrary (non-negative) admissible weights.

If we use the change of variable $\xi = S[u] + 2T[u]\cos\varphi$ in the integrals, we find
\eqn\Tchea{\eqalign{S[u] & = {1 \over \pi}\int_{0}^{\pi} d\varphi\,U(S + 2T[u]\cos\varphi) \ ,\cr T^2[u] & = u + \Big({1 \over \pi}\int_{0}^{\pi} d\varphi\cos\varphi\,U(S[u] + 2T[u]\cos\varphi)\Big)T[u]\ .}}
We then differentiate these relations with respect to $u$, and use an integration by parts to arrive at 
\eqn\Tcheb{\eqalign{\dot{S}[u] & = \dot{S}[u]\,L_0[u] + 2\dot{T}[u]\,L_1[u]\ , \cr 2T[u]\dot{T}[u] & = 1 + 2T[u]\dot{T}[u]\,L_0[u] + T[u]\dot{S}[u]\,L_1[u]\ ,}}
where we have set
\eqn\coefT{\eqalign{L_0[u] & = {1 \over \pi}\int_{0}^{\pi}d\varphi\,U'(S[u] + 2T[u]\cos\varphi)\ , \cr L_1[u] & = {1 \over \pi}\int_{0}^{\pi} d\varphi\cos\varphi\,U'(S[u] + 2T[u]\cos\varphi)\ .}}
Eqs.~\Tcheb\ could have been obtained alternatively upon differentiating
directly
\relRS\ with respect to $u$. The quantities $L_0[u]$ and $L_1[u]$ would 
then be identified as generating functions for paths with height difference $0$ 
or $1$ at their endpoints, and with a marked point
along the path. As such, $L_0[u]$ and $L_1[u]$ are clearly positive. In
our setting, this property follows simply from the expressions 
\eqn\exp{\eqalign{L_0[u] & = \sum_{m \geq 0} {U^{(2m + 1)}(S[u]) \over (2m)!}\,(2T[u])^{2m}\,\Big({1 \over \pi}\int_0^{\pi} d\varphi\cos^{2m}\varphi\Big) \ ,\cr L_1[u] & = \sum_{m \geq 0} {U^{(2m + 2)}(S[u]) \over (2m + 1)!}\,(2T[u])^{2m + 1}\,\Big({1 \over \pi}\int_0^{\pi} d\varphi\cos^{2m + 2}\varphi\Big)\ ,}}
with $T[u] > 0$, while all derivatives of $\xi \mapsto U(\xi)$ at $S[u] > 0$ are
positive since all the $g_k$'s are non-negative (here we eliminate the trivial 
case where only $g_1$ would be non-vanishing). 
Combining the relations \Tcheb\ and eliminating $L_0[u]$, we deduce
\eqn\rfin{(2\dot{T}[u])^2 - (\dot{S}[u])^2 = {\dot{S}[u] \over T[u]\,L_1[u]} > 0\ .}
Since $S[u]$ and $T[u] = \sqrt{R[u]}$ are strictly increasing functions of $u$, the claim follows: $\dot{S}[u] - 2\dot{T}[u] < 0$.

\subsec{Remark on complex valued face weights} 
Let us say a word on more general, maybe complex valued face weights. 
To avoid confusion in this paragraph, we shall denote by ${\cal R}$ and 
${\cal S}$ the power series in $g_k$ satisfying \relRS. We say that a 
sequence of complex numbers $(g_k)_{k\geq 1}$ is {\it sub-admissible} if the 
sequence $(|g_k|)_{k\geq 1}$ is admissible. In this case, ${\cal R}[u]$ and 
${\cal S}[u]$ assume finite values for the sequence $(|g_k|_k)$ and any 
$u \in [0,1]$, that we denote $R_{M}[u]$ and $S_{M}[u]$. Hence, they also 
assume finite values that we denote $R[u]$ and $S[u]$ for the sequence 
$(g_k)_{k\geq 1}$ itself, and we have $|R[u]| \leq R_M[u]$ and 
$|S[u]| \leq S_M[u]$ for all $u$. Moreover, the radius of convergence 
of $V'$ is at least $\gamma_{+,M} = S_{M}[1] + 2\sqrt{R_{M}[1]}$. Recalling 
that $S_M[u] + 2\sqrt{R_M[u]}$ increases with $u$, we have $|S[u] \pm 2\sqrt{R[u]}|
\leq \gamma_{+,M}$. Thus, the disk of convergence of $V'$ contains the 
interior of the discontinuity set of ${\cal W}(x)$ computed with the weights 
$(g_k)_{k\geq 1}$. This ensures also for sub-admissible sequences the validity 
of \Quadrel. The discontinuity set is now:
\eqn\dics{I = \bigcup_{u \in [0,1]} \{x \in {\bf C} \quad (x^2 - 2S[u]x + S^2[u] - 4R[u]) \leq 0\}}
It can be quite complicated, since $R[u]$ and $S[u]$ moves in the complex plane when $u$ increases. Nevertheless, we find:
\eqn\cutoneBB{{\cal W}(x^+) + {\cal W}(x^-) - V'(x) = 0}
for any point $x$ of $I$ such that $|x| < \gamma_{+,M}$ and such that the 
intersection of $I$ with a small ball $B$ centered at $x$ is a piecewise 
${\cal C}^1$ curve separating $B$ in two connected components $B_+$ and 
$B_-$ which both intersect the unbounded component of ${\bf C}\setminus I$. 
$x^{\pm}$ then denote a sequence of points in $B_{\pm}$ which converges 
to $x$. 

\newsec{Conclusion and discussion}

In this paper, we have shown how the gasket decomposition, upon relating 
loop models to models of maps with controlled face degrees, immediately 
provides a functional equation for the resolvent of the model.
We then looked for the largest possible class of models for which this 
functional equation may be solved by the same techniques that were used 
to solve earlier models. In this respect, it was crucial that the cut $[\gamma_-,\gamma_+]$ 
interacts with a single mirror image, which is guaranteed if the 
ring (grand canonical) generating function has a single pole, in which
case a homographic involution naturally enters the game (if the model is
symmetric). For non-negative weights, this involution is moreover
locally decreasing but the solving technique could easily be extended to 
deal with increasing homographic involutions. We discuss in Appendix A how 
a non-generic critical point could emerge in this case.

All the above solvable models (with non-negative weights) can be realized within 
the framework of a single model 
of loops with bending energy, or for twofold avatars of such models in bipartite maps
(such as the twisting loop model). All those models display the same critical behaviours 
as those observed in previously solved $O(n)$ models and no new
universality class is found. In particular, at non-generic critical
points, the critical behaviours are also those of maps with large faces, namely 
the stable maps of \LGM. 

The problem of finding a resolvent satisfying \cutquad\ for 
the quadrangular case with arbitrary $h_1$ and $h_2$ is 
much more involved as the cut now interacts with two mirror cuts 
(its images by $y_+$ and $y_-$). This problem is still open and may 
lead to new interesting critical phenomena. In all generality, 
loop models may involve several poles for the ring generating function
and new techniques are needed to solve these multi-pole models.

As a more promising issue, it is worth mentioning that the nested loop
approach can be generalized to describe Potts model (possibly with defects).
This again leads to a functional equation and it may easily be seen that it
now involves a pair of homographies which are reciprocal of one another, 
rather than a  single involution. This problem is under investigation.

\bigskip
\noindent{\bf Acknowledgments:} We thank P.~Di Francesco for inspiring discussions, and particularly for pointing out the connection with \Lorentz.
The work of G.B. is partly supported
by the ANR project GranMa ``Grandes Matrices Al\'eatoires'' ANR-08-BLAN-0311-01.
J.B. acknowledges the hospitality of Laboratoire d'Informatique Algorithmique: Fondements et Applications (LIAFA) of Universit\'e Paris Diderot and CNRS, where part of this work was completed.

\appendix{A}{Case of a general homographic involution}

We have seen that a model with non-negative local weights cannot lead to a
homographic involution $s(x)$ which is increasing, or for which $s(0) < 0$. 
Let us nevertheless consider here the case of an arbitrary homographic
involution, namely 
\eqn\Sd{s(x) = {\alpha - \beta\, x \over \beta -\delta\, x},\qquad 
\alpha,\beta,\delta \in {\bf R}\ .}
Recall that $s$ has real fixed points iff it is locally decreasing 
over $\bf{R}$, i.e.\ iff $\beta^2 - \alpha\delta > 0$. 
Such a general homographic involution may be obtained in loop models for 
a particular set of real weights. We shall assume here that these weights may be chosen 
in such a way that the resolvent has one cut only on a segment 
$I = [\gamma_-,\gamma_+]$ of the real line. This requirement is not 
an empty condition since the one-cut property is not automatic when 
the weights are not all positive. Note however that, 
under this assumption (and provided the weights are real), the spectral 
density must be positive in the interior of $I$.
\fig{The six situations for the relative positions of $I$ and $s(I)$ (see
text).}{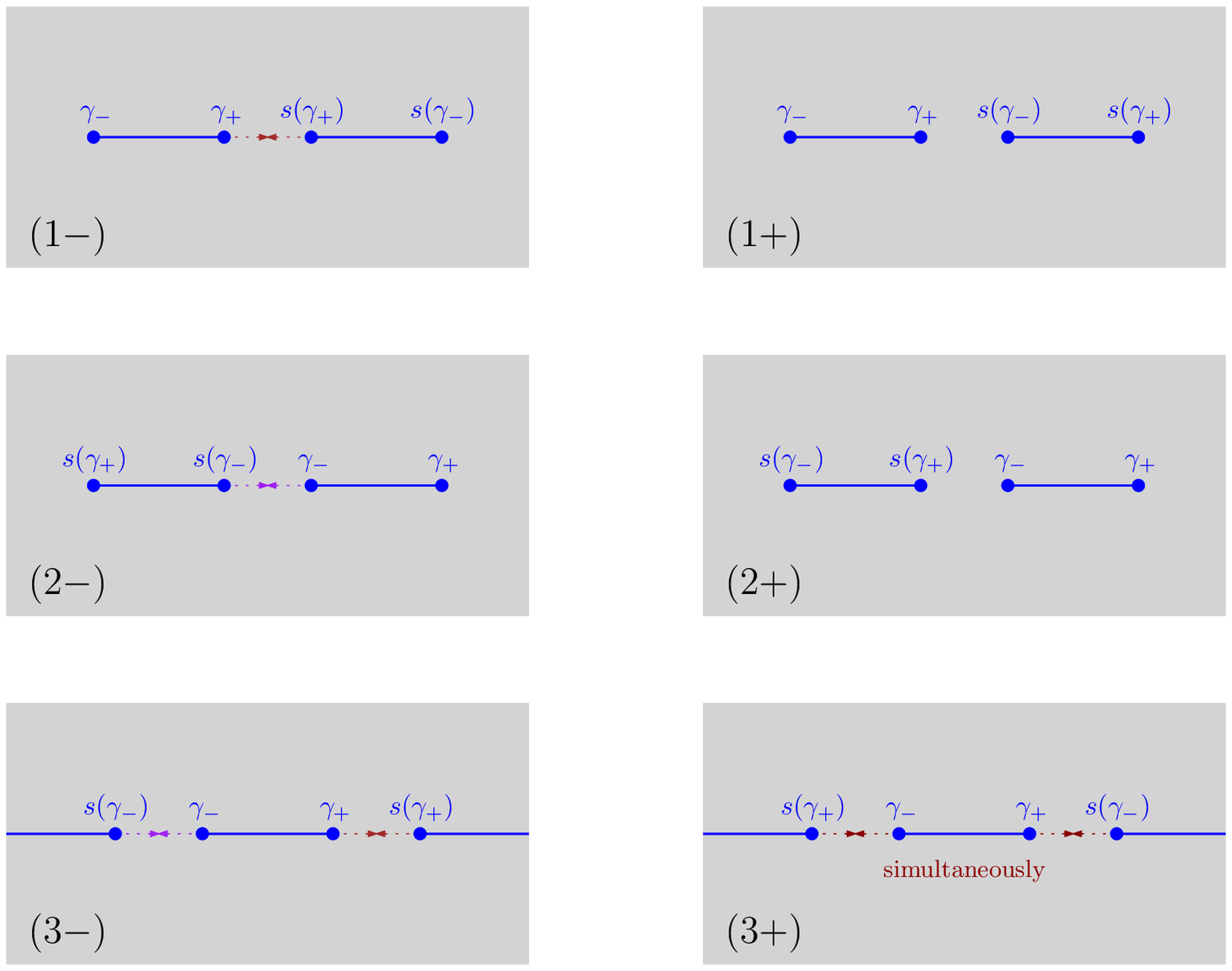}{12.cm}
\figlabel\Schappendix
We may distinguish six situations, displayed in Fig.\Schappendix, 
depending on the local monotonicity of $s$ and on
the relative position of $I$ and $s(I)$. We shall now examine these 
situations and explain when a non-generic critical behaviour can be reached. 

The situation ${\bf 1-}$ corresponds to a locally decreasing involution with 
$s(I)>I$. The non-generic critical behaviour is obtained 
when $\gamma_+$ collides with $s(\gamma_+)$, i.e. is one of the fixed points of $s$.
This case can always be realized in a model with bending energy (see Section 3.2), 
and we have already studied it in details. 

The situation ${\bf 2-}$ corresponds to a locally decreasing involution for which 
$s(I)<I$. The non-generic critical behaviour is obtained when $\gamma_-$ collides with 
$s(\gamma_-)$, i.e. is again one of the fixed points of $s$.

The situation ${\bf 3-}$ corresponds again to a locally decreasing involution 
for which the pole of $s$ belongs to $I$ so that $s(I)$ is split into two semi-infinite
intervals. The non-generic critical behaviour is obtained when $\gamma_+$ collides 
with $s(\gamma_+)$, or when $\gamma_-$ collides with $s(\gamma_-)$, or both. This case also 
arises in a model with bending energy.

The situation ${\bf 1+}$ (resp. ${\bf 2+}$) corresponds to a locally increasing 
involution for which $s(I)>I$ (resp. $s(I)<I$) and does not feature any non-generic 
critical behaviour. Indeed, if $\gamma_{\pm}$ collides with $s(\gamma_{\mp})$, by 
involutivity both extremities of $I$ should collide simultaneously with the extremities of 
$s(I)$, which is clearly not possible. 

The situation ${\bf 3+}$ corresponds to again to a locally increasing involution 
for which the pole of $s$ belongs to $I$, so that $s(I)$ is split into two semi-infinite
intervals. A non-generic behaviour is obtained whenever $\gamma_{+}$ collides with 
$s(\gamma_{-})$, which implies that $\gamma_-$ collides with $s(\gamma_+)$ by involutivity, so that both extremities of $I$ collide simultaneously with the extremities of $s(I)$.

The techniques presented in this article to determine the resolvent $W(x)$ can easily be 
applied to the six situations. The equations of the critical manifold, and in particular 
the expression of $W(x)$ on the non-generic critical manifold in terms of trigonometric 
functions would follow.

\listrefs
\end